# Towards Linear Algebra over Normalized Data

Lingjiao Chen[1]    Arun Kumar[2]    Jeffrey Naughton[3]    Jignesh M. Patel[1]

[1]University of Wisconsin-Madison    [2]University of California, San Diego    [3]Google

[1]{lchen, jignesh}@cs.wisc.edu, [2]arunkk@eng.ucsd.edu, [3]naughton@google.com

## ABSTRACT

Providing machine learning (ML) over relational data is a mainstream requirement for data analytics systems. While almost all ML tools require the input data to be presented as a single table, many datasets are multi-table. This forces data scientists to join those tables first, which often leads to data redundancy and runtime waste. Recent works on "factorized" ML mitigate this issue for a few specific ML algorithms by pushing ML through joins. But their approaches require a *manual* rewrite of ML implementations. Such piecemeal methods create a massive development overhead when extending such ideas to other ML algorithms. In this paper, we show that it is possible to mitigate this overhead by leveraging a popular formal algebra to represent the computations of many ML algorithms: linear algebra. We introduce a new logical data type to represent normalized data and devise a framework of algebraic rewrite rules to convert a large set of linear algebra operations over denormalized data into operations over normalized data. We show how this enables us to *automatically* "factorize" several popular ML algorithms, thus unifying and generalizing several prior works. We prototype our framework in the popular ML environment R and an industrial R-over-RDBMS tool. Experiments with both synthetic and real normalized data show that our framework also yields significant speed-ups, up to 36x on real data.

## 1. INTRODUCTION

The data management industry and academia are working intensively on tools to integrate machine learning (ML) algorithms and frameworks such as R with data platforms [2, 9,10,19,21,24,38]. While almost all ML tools require the input data to be presented as a single table, many datasets are multi-table, typically connected by primary key-foreign key (PK-FK) or more general "M:N" dependencies [33], which forces data scientists to join those tables first. However, such joins often introduce *redundancy* in the data [33], leading to extra storage requirements and runtime inefficiencies due to redundancy in the computations of the ML algorithms.

A few recent works [26, 32, 34, 35] aim to avoid such redundancy by decomposing the computations of some *specific* ML algorithms and pushing them through joins. However, a key limitation of such approaches is that they require *manually* rewriting each ML algorithm's implementation to obtain a "factorized" version. This creates a daunting development overhead in extending the benefits of factorized ML to other ML algorithms. Moreover, the prior approaches are too closely tied to a specific data platform, e.g., an in-memory engine [35] or an RDBMS [26]. This state of the art is illustrated in Figure 1(a) and it raises an important question: *Is it possible to generalize the idea of factorized ML and "automate" its application to a much wider variety of ML algorithms and platforms in a unified manner?*

In this paper, we present the first systematic approach that takes a step towards generalizing and automating factorized ML. Our idea is to use a common formal representation language for ML algorithms: *linear algebra* (LA). Many popular ML algorithms such as linear regression, logistic regression, and K-Means clustering can be expressed succinctly using LA *operators* such as matrix multiplication and inversion [9]. Moreover, data scientists often write new ML algorithms in popular LA-based frameworks such as R [3]. The data management community has embraced LA and R as a key environment for ML workloads [2,4,9]. For example, Oracle R Enterprise (ORE) lets users write LA scripts over an R "DataFrame" that is stored as an in-RDBMS table [2], while Apache SystemML provides an R-like language to scale to data on HDFS [9]. While such systems provide scalability and sophisticated optimizations (e.g., SystemML's hybrid parallelism [8] and SPOOF [17]), they do not optimize LA scripts over normalized data.

Our high-level approach is illustrated in Figure 1(c). Given an ML algorithm in LA (logistic regression in the figure) and the normalized schema, our middleware framework named MORPHEUS *automatically* creates the factorized version of the ML algorithm, i.e., one that operates on the base tables. As illustrated in Figure 1(b), this approach lets us factorize many ML algorithms with one framework, thus mitigating the development overhead. Furthermore, by decoupling how the ML algorithm is factorized from which platform it is run on, MORPHEUS lets us leverage existing scalable LA systems.

Realizing a framework like MORPHEUS is technically challenging due to three crucial desiderata. First is *generality*, i.e., it should be able to handle a wide variety of ML algorithms expressible in LA, as well as both PK-FK and M:N joins. Second is *closure*, i.e., it should ideally rewrite an LA script only into a different LA script so that the internals of the LA system used need not be modified, which could make practical adoption easier. Third is *efficiency*, i.e., it should offer mechanisms to ensure that the factorized version is only used when it is faster than the single-table version.

As a step towards providing high *generality*, in this paper, we focus on a set of LA operations over the data matrix that are common for several popular ML algorithms. These LA operations are listed in Table 1. We introduce a new "logical" data type, the *normalized matrix*, to represent multi-table data inputs in LA. In a sense, our work brings the classical database notion of *logical data independence* [33] to LA systems. To ensure tractability, we focus only on some join schemas that are ubiquitous in practice: "star schema" PK-FK joins and "chain schema" M:N joins. More complex join schemas are left to future work.

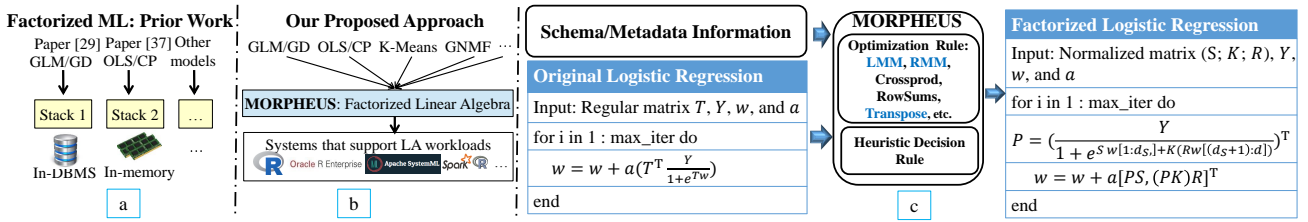

Figure 1: (a) In contrast to prior approaches to factorized ML, which were ML algorithm- and platform-specific, (b) MORPHEUS provides a unified and generic framework to automatically factorize many ML algorithms over any platform that supports LA workloads. (c) Illustration of how MORPHEUS automatically rewrites the standard single-table version of logistic regression into the factorized version using the normalized schema information. The LA operators whose rewrite rules are invoked are highlighted. The heuristic decision rule uses input data statistics to predict if the factorized version will be faster.

To provide *closure*, we devise an extensive framework of *algebraic rewrite rules* that transform an LA operation over a denormalized data matrix into a set of LA operations over the normalized matrix. In a sense, this is a first principles extension of the classical idea of pushing relational operations through joins [12, 37] to LA. Some LA operations such as scalar-matrix multiplication and aggregations are trivial to rewrite and are reminiscent of relational optimization. But more complex LA operations such as matrix-matrix multiplication, matrix cross-product, and matrix inversion enable us to devise novel rewrites that exploit their *LA-specific semantics* with no known counterparts in relational optimization. We also handle matrix transpose. For exposition sake, we describe the rewrite rules for a PK-FK join and then generalize to star schema multi-table PK-FK joins and M:N joins. We apply our framework to four popular and representative ML algorithms to show how they are automatically factorized: logistic regression for classification, linear regression, K-Means clustering, and Gaussian Non-negative Matrix Factorization (GNMF) for feature extraction. Our automatic rewrites largely subsume the ideas in [26, 35] for logistic and linear regression and produce the first known factorized versions of K-Means and GNMF.

Finally, we discuss the *efficiency* trade-offs involved in the rewrites of complex LA operations and present ways to optimize their performance. We also present simple but effective heuristic decision rules to predict when a factorized LA operation might cause slow-downs; this happens in some extreme cases depending on the dimensions of the base tables [26].

We prototype MORPHEUS on standalone R, which is popular for ML-based analytics [3], and the R-over-RDBMS tool Oracle R Enterprise (ORE) [2] (but note that our framework is generic and applicable to other LA systems as well). We present an extensive empirical evaluation using both real and synthetic datasets. Our experiments validate the effectiveness of our rewrite rules and show that MORPHEUS yields speed-ups of up to 36.4x for popular ML algorithms over real data. Compared to a prior ML algorithm-specific factorized ML tool [26], MORPHEUS achieves comparable or higher speed-ups, while offering higher generality. Finally, we also evaluate the scalability of MORPHEUS on ORE.

In summary, this paper makes the following contributions:

- To the best of our knowledge, this is the first paper on generalizing and automating the idea of factorized ML, focusing on ML algorithms expressed in LA.
- We present a mechanism to represent normalized data in LA and an extensive framework of rewrite rules for LA operations. Our framework provides high generality and closure with respect to LA, enabling us to automatically factorize several popular ML algorithms.
- We extend our framework to star schema multi-table PK-FK joins as well as M:N joins.
- We provide prototypes of our framework in both R and ORE and perform an extensive empirical analysis of their performance using both real and synthetic data.

**Outline.** Section 2 presents the problem setup and background. Section 3 introduces the normalized matrix, explains the architecture of MORPHEUS, and dives deep into the rewrite rules. Section 4 applies our framework to four ML algorithms. Section 5 presents the experiments and Section 6 presents the related work. We conclude in Section 7.

## 2. PRELIMINARIES AND BACKGROUND

**Problem Setup and Notation.** For simplicity of exposition, we start with a single PK-FK join. Multi-table joins and M:N joins will be discussed later in Section 3. Consider two tables: $\mathbf{S}(Y, \mathbf{X}_S, K)$ and $\mathbf{R}(\underline{RID}, \mathbf{X}_R)$, where $\mathbf{X}_S$ and $\mathbf{X}_R$ are the *feature vectors*, and $Y$ is called the *target* (for supervised classification and regression). $K$ is the foreign key and $RID$ is the primary key of $\mathbf{R}$. Following [26], we call $\mathbf{S}$ the *entity* table and $\mathbf{R}$ the *attribute* table. The output of the join-project query is denoted by $\mathbf{T}(Y, \mathbf{X}) \leftarrow \pi(\mathbf{S} \bowtie_{K=RID} \mathbf{R})$, wherein $\mathbf{X} \equiv [\mathbf{X}_S, \mathbf{X}_R]$ is the concatenation of the feature vectors. We adopt a standard convention on data representation: let $\mathbf{R}.\mathbf{X}_R$ (resp. $\mathbf{S}.\mathbf{X}_S$, $\mathbf{S}.Y$, $\mathbf{T}.\mathbf{X}$) correspond to the feature matrix $R$ (resp. $S$, $Y$, $T$). Table 2 summarizes our notation.

**Example** (based on [26]). Consider an insurance analyst classifying customers to predict who might churn, i.e., cancel their policy. She builds a logistic regression classifier using a table with customer details: Customers (CustomerID, Churn, Age, Income, EmployerID). EmployerID is the ID of the customer's employer, a foreign key referring to a table about organizations that potentially employ the customers: Employers (EmployerID, Revenue, Country). Thus, $\mathbf{S}$ is Customers, $\mathbf{R}$ is Employers, $K$ is $\mathbf{S}$.EmployerID, RID is $\mathbf{R}$.EmployerID, $\mathbf{X}_S$ is {Age, Income}, $\mathbf{X}_R$ is {Country, Revenue}, and $Y$ is Churn. She joins the tables to bring in $X_R$ because she has a hunch that customers employed by rich corporations in rich countries are unlikely to churn.

Table 1: Operators and functions of linear algebra handled in this paper over a normalized matrix $T$.

| Op Type | Name | Expression | Output Type | Parameter X or x | Factorizable |
|---|---|---|---|---|---|
| Element-wise Scalar Op | Arithmetic Op ($\oslash = +, -, *, /, \hat{} ,$ etc) | $T \oslash x$ or $x \oslash T$ | Normalized Matrix | A scalar | Yes |
| | Transpose | $T^\intercal$ | | N/A | |
| | Scalar Function $f$ (e.g., log, exp, sin) | $f(T)$ | | Parameters for $f$ | |
| Aggregation | Row Summation | rowSums($T$) | Column Vector | N/A | |
| | Column Summation | colSums($T$) | Row Vector | | |
| | Summation | sum($T$) | Scalar | | |
| Multiplication | Left Multiplication | $TX$ | Regular Matrix | $(d_S + d_R) \times d_X$ matrix | |
| | Right Multiplication | $XT$ | | $n_X \times n_S$ matrix | |
| | Cross-product | crossprod($T$) | | N/A | |
| Inversion | Pseudoinverse | ginv($T$) | | N/A | |
| Element-wise Matrix Op | Arithmetic Op ($\oslash = +, -, *, /, \hat{} ,$ etc) | $X \oslash T$ or $T \oslash X$ | | $n_S \times (d_S + d_R)$ matrix | No |

Table 2: Notation used in this paper.

| Symbol | Meaning |
|---|---|
| **R** / $R$ | Attribute table/feature matrix |
| **S** / $S$ | Entity table/feature matrix |
| **T** / $T$ | Join output table/feature matrix |
| $K$ | Indicator matrix for PK-FK join |
| $I_S$ / $I_R$ | Indicator matrices for M:N join |
| $Y$ | Target matrix (regression and classification) |
| $n_R$ / $n_S$ | Number of rows in R / S (and T) |
| $d_R$ / $d_R$ | Number of columns in R / S |
| $d$ | Total number of features, $d_S + d_R$ |
| $n_U$ | M:N join attribute domain size |

**Linear Algebra (LA) Systems and R.** LA is an elegant formal language in which one can express many ML algorithms [9, 16]. Data are formally represented as matrices, with scalars and vectors being special cases. LA operators map matrices to matrices. Basic operators include unary operators such as element-wise exponentiation, as well as binary operators such as matrix-matrix multiplication. Derived operators include Gram matrix and aggregation operators. An LA system is a system that supports matrices as a first class data type, as well as elementary and derived LA operators such as indexing, matrix multiplication, and pseudo-inverse. Widely used examples include R, Matlab, SAS, and Python's NumPy. In particular, open source R has gained immense popularity and has free ML libraries for various domains [3]. R is primarily an in-memory tool but recent systems built by the data management community enables one to scale LA scripts written in R (or R-like languages) to data resident in an RDBMS, Hive/Hadoop, and Spark. Examples of such "R-based analytics systems" include RIOT-DB [38], Oracle R Enterprise (ORE), Apache SystemML [9], and SparkR [4]. In this paper, we implement our framework on standard R and also ORE. Note that our ideas are generic enough to be applicable to other LA systems such as Matlab, NumPy, other R-based analytics systems, or TensorFlow as well.

**Factorized ML.** Factorized ML techniques were studied in a recent line of work for a few specific ML algorithms [26, 34, 35]. We briefly explain a key representative technique that introduced this paradigm: "factorized learning" from [26]. Given a model vector $w$, a GLM with gradient descent computes the inner products $w^\intercal x$, in each iteration, for each feature vector $x$ from **T**. Since **T** has redundancy across tuples, this multiplication involves redundant computations across tuples, which is what factorized learning avoids. The crux of its idea is to decompose the inner products over $x$ into inner products over the feature vectors $x_S$ and $x_R$ from the two base tables. Thus, the partial inner products from **R** can be saved and then reused for each tuple in **S** that refers to the same tuple in **R**. This is correct because $w^\intercal x = w_S^\intercal x_S + w_R^\intercal x_R$, wherein $w_S$ (resp. $w_R$) is the projection of $w$ to the features from **S** (resp. **R**). Figure 1(c) illustrates how logistic regression is factorized. Factorized learning often has significantly faster runtimes.

**Problem Statement and Scope.** We ask: *Is it possible to transparently "factorize" a large set of LA operations over T that are common in ML into operations over S and R without losing efficiency?* Our goal is to devise an integrated framework of such algebraic rewrite rules for the key application of automatically "factorizing" ML algorithms written in LA, which could mean that developers need not manually rewrite ML implementations from scratch. The challenge in devising such a framework is in preserving *generality* (i.e., applicability to many ML algorithms and both PK-FK and M:N joins), *closure* (i.e, rewrites only produce a different LA script), and *efficiency* (i.e., faster or similar runtimes). Most LA systems support a wide variety of operations on matrices. For tractability sake, we restrict our focus to a large subset of LA operations that still support a wide variety of ML algorithms; Table 1 lists and explains these operations.

## 3. FACTORIZED LINEAR ALGEBRA

We introduce the normalized matrix, give an overview of how MORPHEUS is implemented, and dive deep into our framework of rewrite rules for a single PK-FK join. We then extend our framework to multi-table joins and M:N joins.

## 3.1 The Normalized Matrix

We introduce a new multi-matrix logical data type called the *normalized matrix* to represent normalized data. It is called a logical data type because it only layers a logical abstraction on top of existing data types. For simplicity of exposition, this subsection focuses on a PK-FK join; Section 3.5 and 3.6 present the extensions of the normalized matrix to star schema multi-table PK-FK joins and M:N joins, respectively. Note that each **R**.$RID$ in the attribute table **R** can be mapped to its sequential row number in the matrix $R$. Thus, **S**.$K$ can be viewed as an attribute containing entries that are the row numbers of $R$. An *indicator* matrix $K$ of size $n_S \times n_R$ can thus be constructed as follows:

$$K[i,j] = \begin{cases} 1, & if\ i^{th}\ row\ of\ \mathbf{S}.K = j \\ 0, & otherwise \end{cases}$$

The *normalized matrix* corresponding to $T$ is defined as a matrix triple $T_N \equiv (S, K, R)$. One can verify that $T = [S, KR]$.[1] It is worth noting that $K$ is a highly *sparse* matrix. In fact, the PK-FK relationship implies that the number of non-zero elements in each row of $K$ is 1. Thus, $nnz(K)$, the number of non-zero elements in $K$, is exactly $n_S$. Without loss of generality, assume $\forall j, nnz(K[,j]) > 0$, i.e., each tuple in **R** is referred to by at least one tuple in **S**. Otherwise, we can remove from **R** all the tuples that are never referred to in **S**. Note that any of $R$, $S$, and $T$ can be dense or sparse.

A natural question is how expressive our abstraction is for ML, i.e., what kind of ML algorithms it can benefit efficiency-wise. In short, ML algorithms whose data-intensive computations can be *vectorized* as elementary or derived LA operations over the feature matrix $T$ in bulk can benefit from our abstraction. This is expected because vectorized computation is a key efficiency assumption made by almost all LA systems, including R, Matlab, and SystemML [9]. Since our work builds on top of such LA systems, the same assumption carries over to our work. In practice, the data-intensive computations of many popular ML algorithms, including supervised ML, unsupervised ML, and feature extraction algorithms can be vectorized, as we illustrate in detail in Section 4; [9] also provides more examples.[2]

## 3.2 Overview of Morpheus

MORPHEUS is an implementation of the normalized matrix and our framework of rewrite rules as a class in standard R and ORE. Our class has three matrices: $S$, $K$, and $R$. All LA operators in Table 1 are overloaded to support our class. The details of how the operators are executed over normalized matrices are the subject of Section 3.3. Interestingly, some operators output a normalized matrix, which enables MORPHEUS to propagate the avoidance of data redundancy in a given LA script with multiple operators. Another interesting point is how to handle the transpose of a normalized matrix. A straightforward way is to create a new class for transposed normalized matrices and overload the operators again. Instead, we adopt a different approach that makes our implementation more succinct and exploits more rewrite opportunities. We add a special binary "flag" to indicate if a normalized matrix is transposed. If the flag is false, Section 3.3 rules are used; otherwise, we use the rewrite rules for transpose presented in the appendix. Compared to the straightforward approach, our approach avoids computing repeated transposes and allows developers to focus on only one new class.[3]

Finally, we explain how to construct a normalized matrix from the base tables **S** and **R** given as, say, CSV files. We illustrate this process with a code snippet. For the sake of brevity, we assume that $RID$ and $K$ are already sequential row numbers. Note that "list" is used to allow different data types (e.g., dense or sparse) and multi-table data.

```
S = read.csv("S.csv")          //foreign key name K
R = read.csv("R.csv")
K = sparseMatrix(i=1:nrow(S), j=S[,"K"], x=1)
TN = NormalizedMatrix(EntTable=list(S),
        AttTables=list(R), KIndicators=list(K))
```

Overall, MORPHEUS is packaged as easy-to-use libraries for both standard R and Oracle R Enterprise. Our code has been open sourced on the project webpage: http://cseweb.ucsd.edu/~arunkk/morpheus.

## 3.3 Factorized Linear Algebra Operators

We now dive deep into our framework of algebraic rewrite rules for the groups of operators listed in Table 1.

### 3.3.1 Element-wise Scalar Operators

These are trivial to rewrite but they are ubiquitous in ML. They include multiplication and addition of a matrix with a scalar, element-wise exponentiation, and element-wise scalar functions, e.g., $log$ and $exp$. The output is a normalized matrix with the same structure as the input. The rewrite rules are given below; "⊘" represents a binary arithmetic operator, $x$ is a scalar, and $f$ is a scalar function.

$$T \oslash x \rightarrow (S \oslash x, K, R \oslash x) \ ; \ x \oslash T \rightarrow (x \oslash S, K, x \oslash R)$$
$$f(T) \rightarrow \ (f(S), K, f(R))$$

In the above, $T \oslash x \rightarrow (S \oslash x, K, R \oslash x)$ means that an operation $T \oslash x$ can be replaced implicitly with operations on the normalized matrix $(S, K, R)$ to yield a new normalized matrix $(S \oslash x, K, R \oslash x)$. These rewrites avoid redundant computations. For instance, computing $3 \times T$ requires $n_S(d_S + d_R)$ multiplications but computing $(3 \times S, K, 3 \times R)$ requires only $n_S d_S + n_R d_R$. The ratio of these two quantities is the ratio of the size of $T$ to the total size of $S$ and $R$. The speed-ups depend on this ratio and thus, the speed-ups could be significant when this ratio is large.

### 3.3.2 Aggregation Operators

These include rowSums($T$), which sums the matrix row-wise, colSums($T$), which sums the matrix column-wise, and sum($T$), which adds up all of the elements. These operators also arise frequently in ML, especially when computing loss or gradient values, which are aggregates over examples (or features). The rewrite rules are as follows.

---

[1] For ease of exposition, we abuse the notation slightly and use $T$ itself for $T_N$ when it is clear from the context that the rewrites operate over the normalized matrix rather than the regular matrix, e.g., in Table 1 and Section 3.3.

[2] But not all ML algorithms are amenable to vectorized bulk computations over $T$, e.g., stochastic gradient descent (SGD) will likely be a poor fit for LA systems, since it updates the model after each example or mini-batch from $T$ [19]. We leave a deeper study of SGD to future work.

[3] Our architecture fits easily into any interpreted environments for LA; we leave to future work an integration with a compiler environment such as SystemML [9].

Figure 2: Illustration of factorized LMM. (A) Materialized LMM $TX$. (B) The first step in factorized LMM is $SX[1:d_S,] = Z_1$ (say). Note that $d_S = 2$. (C) Next, $RX[d_S + 1 : d,] = Z_2$ (say). Note that $d = 4$. (D) Then, $KZ_2 = Z_3$ (say). Finally, the factorized LMM is $Z_1 + Z_3$, the same as the result in (A).

$$\text{rowSums}(T) \rightarrow \text{rowSums}(S) + K\text{rowSums}(R)$$
$$\text{colSums}(T) \rightarrow [\text{colSums}(S), \text{colSums}(K)R]$$
$$\text{sum}(T) \rightarrow \text{sum}(S) + \text{colSums}(K)\text{rowSums}(R)$$

The rule for rowSums pushes down the operator to before the join and then multiplies the pre-aggregated $R$ with $K$, before adding both parts. The rule for colSums, however, first pre-aggregates $K$ before multiplying it with $R$ and then attaches it to the pre-aggregated $S$. Finally, the rewrite rule for sum is more complicated and involves a sum push-down along with a rowSums and a colSums. These rewrite rules are essentially the LA counterparts of SQL aggregate push-down optimizations in RDBMSs [12, 37]. By extending such operator rewrite ideas to LA operations, our work makes them more widely applicable, especially, for LA-based ML workloads that may not use an RDBMS.

### 3.3.3 Left Matrix Multiplication (LMM)

LMM is an important and time-consuming operator arising in many ML algorithms, typically for multiplying the data matrix with a model/weight vector. In fact, it arises in all of GLMs, K-Means, and GNMF. Interestingly, a special case of LMM is the key operation factorized in [26]. Our rewrite rule expresses that idea in LA and generalizes it to a weight matrix, not just a weight vector. The rewrite rule is as follows; $X$ is a regular $d \times d_X$ ($d_X \geq 1$) matrix.

$$TX \rightarrow SX[1:d_S,] + K(RX[d_S + 1 : d,])$$

Essentially, we first split up $X$, then pre-multiply with $S$ and $R$ separately, and finally add them. A subtle but crucial issue is the order of the multiplication in the second component. There are two orders: (1) $(KR)X[d_S + 1 : d_S + d_R,]$, and (2) $K(RX[d_S + 1 : d,])$. The first is equivalent to materializing (a part of) the output of the join, which causes computational redundancy! The second avoids the computational redundancy and thus, we use the second order. Most LA systems, including R, allow us to fix the multiplication order using parentheses. A key difference with [26] is that their approach stores the partial results over $R$ in an in-memory associative array. We avoid using associative arrays, which are not a native part of most LA systems, and instead, use regular matrix multiplications. While this could lead to a small performance penalty, it enables us to satisfy the *closure* property explained before. Figure 2 illustrates how factorized LMM works.

### 3.3.4 Right Matrix Multiplication (RMM)

RMM also appears in many ML algorithms, including GLMs, especially when the normalized matrix is transposed (transposed operators are discussed in detail in our technical report [13]). Let $X$ be a regular $m \times n_S$ ($m \geq 1$) matrix. The rewrite rule is as follows.

$$XT \rightarrow [XS, (XK)R]$$

This rewrite does not need to split up $X$ but pushes down the RMM to the base tables and then attaches the resultant matrices. Once again, the second component has two possible orders, with the one that is not used being logically equivalent to materializing the join output.

A related matrix multiplication operation involves multiplying two normalized matrices; we call this operation Double Matrix Multiplication (DMM). In contrast to LMM and RMM, to the best of our knowledge, DMM does not arise in any popular ML algorithm. Nevertheless, we show in the appendix that it is indeed possible to rewrite even a DMM into operations over the base tables' matrices although the rewrite is more complicated.

### 3.3.5 Cross-product

The cross-product of a matrix $T$, denoted $\text{crossprod}(T)$ in R, is equivalent to $T^\intercal T$.[4] Most LA systems offer cross-product as a unary function.[5] It arises in ML algorithms where feature-feature interactions are needed, e.g., linear regression using normal equations, covariance, and PCA [16]. Interestingly, alternative rewrites are possible for crossprod. We start with a straightforward "naive method" in Algorithm 1. Since $T^\intercal T$ is symmetric, we need only half of the output matrix and its diagonal. Thus, this rewrite first computes the lower-left (and upper-right) by multiplying $R^\intercal$ with the product of $K^\intercal$ and $S$, which avoids materialization. Second, it computes the cross-product of $S$ for the upper-left. Third, it computes the cross-product of $K$ and thus, the cross-product of $KR$ without materializing the join. Finally, the results are stitched appropriately. The approach in [35] to factorize a part of the so-called "co-factor" matrix for linear regression is similar.

While already a bit optimized, Algorithm 1 still has two inefficiency issues. First, it does not fully exploit the symmetry of some matrices. Second, transposed multiplication of a sparse matrix ($K^\intercal K$) is a non-trivial cost in many cases. We

---
[4] By convention, data examples are rows in R, which means crossprod is actually the Gram matrix in LA textbooks [22].
[5] There is also a binary version: $\text{crossprod}(T_1, T_2) = T_1^\intercal T_2$. If only $T_2$ is normalized, it is RMM; if only $T_1$ is normalized, it is transposed RMM, which is discussed in Section 3.4. If both are normalized, it is a transposed double multiplication, which is discussed in the appendix.

**Algorithm 1:** Cross-product (Naive method)

$P = R^\intercal(K^\intercal S)$
**return** $\begin{bmatrix} S^\intercal S & P^\intercal \\ P & R^\intercal((K^\intercal K)R) \end{bmatrix}$

**Algorithm 2:** Cross-product (Efficient method)

$P = R^\intercal(K^\intercal S)$
**return**
$\begin{bmatrix} \text{crossprod}(S) & P^\intercal \\ P & \text{crossprod}\left((\text{diag}(\text{colSums}(K)))^{\frac{1}{2}} R\right) \end{bmatrix}$

present a novel rewrite–the "efficient method"–that resolves both issues. The first one is resolved by using crossprod($S$) directly instead of $S^\intercal S$. This reduces about $\frac{1}{2}n_S d_S^2$ arithmetic computations. The second one is more subtle; we make three observations: (1) $K^\intercal K$, denoted $K_p$, is not only symmetric but also *diagonal*. (2) $K_p[i,i]$ is the number of ones in the $i^{th}$ column of $K$. Thus, $K_p \equiv \text{diag}(\text{colSums}(K))$, where diag is a standard LA operation that creates a diagonal matrix given a vector. Third, denoting the element-wise square root of $K_p$ by $K_p^{\frac{1}{2}}$, we have:

$$R^\intercal(K^\intercal K)R = \text{crossprod}(K_p^{\frac{1}{2}} R)$$

Compared to the expression on the left, the one on the right avoids transposing a sparse matrix and replaces several matrix multiplications with a single cross-product. Hence, about $\frac{1}{2}n_R d_R^2$ arithmetic computations will be saved by the expression on the right. Integrating these observations, the efficient method is presented in Algorithm 2.

### 3.3.6 Matrix Inversion Operators

Note that $T$ is seldom a square matrix in practice; hence, it will typically not be directly invertible. Interestingly, we show in the appendix that, even if $T$ is actually a square matrix, it is highly likely to be singular. This is because non-singularity imposes a strict constraint on the relative dimensions of the base tables [13]. Thus, we consider *ginv*, the Moore-Penrose pseudo-inverse and provide the rewrites rules for it below. The rewrite rules for *solve*, which is often used to avoid a full inversion, are similar.

$$ginv(T) \rightarrow ginv(\text{crossprod}(T))T^\intercal, \text{ if } d < n$$
$$ginv(T) \rightarrow T^\intercal ginv(\text{crossprod}(T^\intercal)), \text{ o/w}$$

### 3.3.7 Non-Factorizable Operators

Element-wise matrix arithmetic operators such as matrix addition do not necessarily have redundancy introduced into their computations by joins. Thus, we call such operators "non-factorizable." To see why such operators may not have redundancy, consider the matrix addition $T + X$, where $T$ is the normalized matrix and $X$ is a regular matrix of the same size, i.e., $n_S \times (d_S + d_R)$. In general, it is possible that $X$ has no redundancy, i.e., all its entries are unique, say, $X[i,j] = ((i-1)n_S + j)n_S(d_S + d_R)$. Now, suppose that all entries in $S$ and $R$ are just 1, which means all entries of $T$ are also just 1. Thus, $T$ has a large amount of redundancy. But $T + X$ simply adds 1 to each element of $X$. Since the elements of $X$ are all unique, there is no redundancy in this computation, which is why we say $T + X$ is non-factorizable. In general, there could be "instance-specific" redundancy in $X$, e.g., some elements just happen to be repeated by chance. Exploiting such instance-specific redundancy is beyond the scope of this work. Fortunately, element-wise matrix arithmetic operations are rare in ML; to the best of our knowledge, there is no popular ML algorithm where these operations are the runtime bottleneck.[6] Thus, we ignore these operations henceforth.

### 3.4 Runtime Complexity Analysis

We now present the runtime complexity of the factorized LA operators yielded by our rewrite rules. Instead of just the "big O" notation, we provide the proportional dependency of the number of arithmetic computations (multiplications and additions) in terms of the dimensions of the base table's matrices. Table 3 presents the expressions for the standard (materialized) and factorized versions. Due to space constraints, we discuss these expressions in more detail (and also provide the more tedious expressions for $ginv$) in the appendix. To understand the asymptotic speed-ups of the factorized versions over the corresponding standard versions, let $TR \equiv \frac{n_S}{n_R}$ and $FR \equiv \frac{d_R}{d_S}$ denote the tuple ratio and feature ratio, respectively. For most of the LA operators, the speed-ups converge to $1 + FR$ (resp. $TR$) as $TR$ (resp. $FR$) goes to infinity. The speedup for crossprod, however, converges to $(1 + FR)^2$ as $TR$ increases, since its runtime complexity is quadratic in $d$.

Table 3: Arithmetic computations of the standard algorithms and factorized ones. Lower order terms are ignored.

| Operator | Standard | Factorized |
|---|---|---|
| Scalar Op | $n_S(d_S + d_R)$ | $n_S d_S + n_R d_R$ |
| Aggregation | | |
| LMM | $d_X n_S(d_S + d_R)$ | $d_X(n_S d_S + n_R d_R)$ |
| RMM | $n_X n_S(d_S + d_R)$ | $n_X(n_S d_S + n_R d_R)$ |
| crossprod | $\frac{1}{2}(d_S + d_R)^2 n_S$ | $\frac{1}{2}d_S^2 n_S + \frac{1}{2}d_R^2 n_R + d_S d_R n_R$ |

### 3.5 Extension to Multi-table Joins

We now extend our framework to multi-table PK-FK joins, specifically, star schema joins, which are ubiquitous in practice. For example, in recommendation systems such as Netflix and Amazon, the table with ratings has two foreign keys referring to tables about users and products. Thus, there is one entity table and two attribute tables. Formally, the schema is as follows: one entity table/matrix $S$, $q$ attribute tables, $R_1, \ldots, R_q$, and $q$ associated PK-FK matrices $K_1, \ldots, K_q$. The materialized join output $T$ is $[S, K_1 R_1, \ldots, K_q R_q]$. The extended normalized matrix is the tuple $(S, \ldots K_1, K_2, \ldots, K_q, R_1, R_2, \ldots, R_q)$. We now present the extended rewrite rules.

---

[6]Such operations may arise in non-ML applications of LA, e.g., scientific simulations and financial engineering, but it is not clear if these application have normalized data.

**Element-wise Scalar Operators.** The extension is straightforward and as follows.

$$T \oslash x \to (S \oslash x, K_1, \ldots, K_q, R_1 \oslash x, \ldots, R_q \oslash x)$$
$$x \oslash T \to (x \oslash S, K_1, \ldots, K_q, x \oslash R_1, \ldots, x \oslash R_q),$$
$$f(T) \to (f(S), K_1, \ldots, K_q, f(R_1), \ldots, f(R_q)).$$

**Aggregation Operators.** These require pre-aggregation of each $R_i$ using $K_i$ and then combining the partial results, shown as follows.

$$\text{colSums}(T) \to [\text{colSums}(S), \text{colSums}(K_1)R_1,$$
$$\cdots, \text{colSums}(K_q)R_q]$$
$$\text{rowSums}(T) \to \text{rowSums}(S) + \sum_{i=1}^{q} K_i \text{rowSums}(R_i)$$
$$\text{sum}(T) \to \text{sum}(S) + \sum_{i=1}^{q} \text{colSums}(K_i)\text{rowSums}(R_i)$$

**LMM.** We need some notation. Let the dimensions of $R_i$ be $n_{Ri} \times d_{Ri}$. Thus $d = d_S + \sum_{i=1}^{Q} d_{Ri}$. Define $d'_i = d_S + \sum_{j=1}^{i} d_{Rj}$, for $i = 1$ to $q$, and $d'_0 = d_S$. Given $X$ of size $d \times m$ ($m \geq 1$), the rewrite is as follows.

$$TX \to SX[1:d_S,] + \sum_{i=1}^{q} K_i(R_i X[d'_{i-1}+1:d'_i,])$$

**RMM.** Note that the dimensions of $K_i$ is $n_S \times n_{Ri}$. Given $X$ of size $m \times n_S$ ($m \geq 1$), the rewrite is as follows.

$$XT \to [XS, (XK_1)R_1, \ldots, (XK_q)R_q]$$

**Cross-product.** For the sake of readability, let $K_{R_i}$ and $\text{cp}(X)$ denote $K_i R_i$ and $\text{crossprod}(X)$, respectively. Using the block matrix multiplication, $\text{cp}(T) = T^\intercal T = [S, K_1 R_1, \ldots, K_q R_q]^\intercal [S, K_1 R_1, \ldots, K_q R_q]$ can be rewritten as follows.

$$\begin{bmatrix} \text{cp}(S) & S^\intercal K_{R_1} & S^\intercal K_{R_2} & \cdots & S^\intercal K_{R_q} \\ K_{R_1}^\intercal S & \text{cp}(K_{R_1}) & K_{R_1}^T K_{R_2} & \cdots & K_{R_1}^T K_{R_q} \\ K_{R_2}^\intercal S & K_{R_2}^\intercal K_{R_1} & \text{cp}(K_{R_2}) & \cdots & K_{R_2}^\intercal K_{R_q} \\ \vdots & \vdots & \vdots & \ddots & \vdots \\ K_{R_q}^\intercal S & K_{R_q}^\intercal K_{R_1} & K_{R_q}^\intercal K_{R_2} & \cdots & \text{cp}(K_{R_q}) \end{bmatrix}$$

Since $\text{cp}(T)$ is symmetric. we only need to compute the upper right parts of it, i.e., all diagonal block matrices, $S^\intercal K_{R_i}$, and $K_{R_i}^\intercal K_{R_j}$. For each diagonal block matrix $\text{cp}(K_{R_i})$, we use the rewrite rule $\text{crossprod}\left(\left(\text{diag}\left(\text{colSums}(K_i)\right)\right)^{\frac{1}{2}} R_i\right)$. For $S^\intercal K_{R_i}$ and $K_{R_i}^\intercal K_{R_j}$, $(S^\intercal K_i)R_i$ and $R_i(K_i^\intercal K_j)R_j$ are used, respectively.

## 3.6 Extension to M:N Joins

We now briefly explain how our framework can be extended to handle a general non-PK-FK equi-join ("M:N" join) between **S** (or a projection of it that excludes $Y$) and **R**. We discuss the case of multi-table M:N joins in the appendix. Let the join attribute in **S** (resp. **R**) be denoted $J_S$ (resp. $J_R$). Attach attributes $N_S$ and $N_R$ to the corresponding tables to encode row numbers, i.e., $N_S$ (resp. $N_R$) takes values from 1 to $n_S$ (resp. $n_R$). We need to capture which tuples (rows) of **S** and **R** get mapped to which rows of $\mathbf{T} = \mathbf{S} \bowtie_{J_S = J_R} \mathbf{R}$. To do so, first compute $\mathbf{T}' = \pi_{N_S, J_S}(\mathbf{S}) \bowtie_{J_S = J_R} \pi_{N_R, J_R}(\mathbf{R})$ with non-deduplicating projections (potentially, a relational cross-product of the projected join columns). Then, create two indicator matrices $I_S$ and $I_R$ of dimensions $|\mathbf{T}'| \times n_S$ and $|\mathbf{T}'| \times n_R$ respectively:

$$[I_S|I_R]([i,j]) = \begin{cases} 1, & \text{if } i^{th} \text{ row of } \mathbf{T}'.[N_S|N_R]) = j \\ 0, & \text{otherwise} \end{cases}$$

$I_S$ and $I_R$ are also very sparse: $nnz(I_S) = nnz(I_R) = |\mathbf{T}'|$. Without loss of generality, assume each column of $I_S$ and $I_R$ has at least one non-zero, i.e., each tuple of **S** and **R** contributes to at least one tuple of **T**; otherwise, we can remove those tuples a priori. The extended normalized matrix is now $(S, I_S, I_R, R)$ and it is clear that $T = [I_S S, I_R R]$. The extensions to our implementation (Section 3.2) are now straightforward. For brevity, we skip the modified rewrite rules here and present them in the appendix.

## 3.7 Will Rewrites Always Be Faster?

Our rewrites avoid computational redundancy caused by joins.[7] But if the joins are (too) selective and/or introduce no redundancy, the rewrites could worsen performance because $T$ could become smaller than $S$ and $R$ put together. This dichotomy is an instance of the classical problem of cardinality estimation; it is orthogonal to our work and we leave it to future work to integrate sophisticated cardinality estimation ideas into LA systems. In this work, We drop tuples of **S** and **R** that do not contribute to **T**, as explained in Section 3.1 and 3.6. Since many ML algorithms are iterative, this pre-processing time is relatively minor.[8]

But interestingly, in some extreme cases, even after such pre-processing and even if the joins introduce some redundancy, rewrites could worsen performance because the overheads caused by the extra LA operations could dominate the computational redundancy saved. Empirically (Section 5.1), we found such slow-downs to be almost always < 2x, but it is still helpful to predict and avoid these. Using runtime "cost models" for LA operators, say, based on BLAS [30] is one option. However, this ties us too much to a specific LA system back-end and violates genericity, while also imposing the burden of system- and machine-specific cost calibrations on the user (CPU clock frequency, cache sizes, etc.). Thus, we consider a simpler system-agnostic approach that does not need cost models for the operators; instead, we use a simple *heuristic decision rule* that thresholds on the tuple ratio and feature ratio (explained in Section 3.4) to predict if the redundancy saved by the factorized version will be substantial enough. The thresholds are set conservatively. We explain more about why our approach is feasible and what our decision rule looks like in Section 5.1.

## 4. APPLICATION TO ML ALGORITHMS

We now show how MORPHEUS automatically "factorizes" a few popular ML algorithms. We pick a diverse and representative set of ML algorithms: logistic regression for classification, least squares for regression, K-Means for clustering, and Gaussian non-negative matrix factorization (GNMF) for feature extraction. For each algorithm, we present the standard single-table version of their LA scripts, followed by the "factorized" versions for a PK-FK join. Note that these can be easily extended to multi-table joins and M:N joins using rewrite rules from Sections 3.5 and 3.6, respectively. These rewrites are shown for illustration only; MORPHEUS uses the rewrite rules on-the-fly without code regeneration.

---

[7]Our rewrites do not alter the outputs of the operators, assuming exact arithmetic, which is standard for rewrite optimizations in the LA systems literature [9, 38]. We leave a numerical analysis for finite-precision arithmetic to future work. Empirically, we saw that ML accuracy was unaffected.
[8]We verified this empirically (see appendix).

**Logistic Regression for Classification.** Algorithm 3 presents the standard algorithm using gradient descent (GD); the automatically factorized version is in Algorithm 4. The following rewrite rules are used: LMM for $Tw$ and transposed LMM (explained in the appendix) for $T^\intercal P^\intercal$.

---

**Algorithm 3:** Logistic Regression (Standard)

**Input:** Regular matrix $T, Y, w, \alpha$
**for** $i$ $in$ $1 : max\_iter$ **do**
  | $w = w + \alpha * (T^\intercal(Y/(1 + \exp(Tw))))$
**end**

---

**Algorithm 4:** Logistic Regression (Factorized)

**Input:** Normalized matrix $(S, K, R), Y, w, \alpha$
**for** $i$ $in$ $1 : max\_iter$ **do**
  $P = (Y/(1 + \exp(Sw[1 : d_S,] +$
  $\qquad\qquad K(Rw[d_S + 1 : d_S + d_R,]))))^\intercal$
  $w = w + \alpha * [PS, (PK)R]^\intercal$
**end**

---

**Least Squares Linear Regression.** Algorithm 5 presents the standard algorithm using the normal equations; Algorithm 6 presents the factorized version. The following rewrite rules are used: cross-product for crossprod($T$) and transposed LMM for $T^\intercal Y$. If $d$ is too large, or if the cross-product is singular, GD is used instead; this is similar to Algorithm 3 and Algorithm 4 and for brevity sake, we skip it here and present it in the appendix. A hybrid algorithm that constructs the so-called "co-factor" matrix (using the cross-product) and then uses GD was presented and factorized in [35]. Their algorithm can also be automatically factorized by MORPHEUS; we discuss this in more detail in the appendix.

---

**Algorithm 5:** Linear Regression (Standard)

**Input:** Regular matrix $T, Y, w$
$w = ginv(\text{crossprod}(T))(T^\intercal Y)$

---

**Algorithm 6:** Linear Regression (Factorized)

**Input:** Normalized matrix $(S, K, R), Y, w$
$P = ginv(\text{crossprod}((S, K, R)))$  //Use Algo. 2
$w = P([Y^\intercal S, (Y^\intercal K)R])^\intercal$

---

**K-Means Clustering.** The factorized version is in Algorithm 7; the standard version is presented in the appendix due to space constraints. The following rewrite rules are used: element-wise exponentiation and aggregation for rowSums($T\hat{}2$), LMM for $TC$, and transposed LMM for $T^\intercal A$. Note that K-Means requires matrix-matrix multiplications, not just matrix-vector multiplications. This demonstrates a key benefit of the generality of our approach.

**GNMF for Feature Extraction.** Algorithm 8 presents the factorized version; the standard version is presented in the appendix due to space constraints. The following rewrite rules are used: RMM and LMM for $W^\intercal T$ and $TH$ respectively. Similar to K-Means, GNMF also requires full matrix-matrix multiplications.

---

**Algorithm 7:** K-Means Clustering (Factorized)

**Input:** Normalized matrix $(S, K, R)$, # centroids $k$
//Initialize centroids matrix $C \in \mathbb{R}^{d \times k}$
//$\mathbf{1}_{a \times b}$ represents an all 1 matrix in $\mathbb{R}^{a \times b}$; used for replicating a vector row-wise or column-wise
//1. Pre-compute $l^2$-norm of points for distances
$D_T = (\text{rowSums}(S\hat{}2) + K\text{rowSums}(R\hat{}2))\mathbf{1}_{1 \times k}$
$S_2 = 2 \times S; R_2 = 2 \times R$
**for** $i$ $in$ $1 : max\_iter$ **do**
  //2. Compute pairwise squared distances; $D^{n \times k}$ has
  points on rows and centroids/clusters on columns.
  $D = D_T + \mathbf{1}_{n \times 1}\text{colSums}(C\hat{}2) - (S_2C + K(R_2C))$
  //3. Assign each point to nearest centroid; $A^{n \times k}$ is
  a boolean (0/1) assignment matrix
  $A = (D == (\text{rowMin}(D)\mathbf{1}_{1 \times k}))$
  //4. Compute new centroids; denominator counts
  number of points in the new clusters, while
  numerator adds up assigned points per cluster
  $C = [A^\intercal S, (A^\intercal K)R]^\intercal/(\mathbf{1}_{d \times 1}\text{colSums}(A))$
**end**

---

**Algorithm 8:** Gaussian NMF (Factorized)

**Input:** Normalized matrix $(S, K, R)$, rank $r$
//Initialize $W$ and $H$
**for** $i$ $in$ $1 : max\_iter$ **do**
  $P = [W^\intercal S, (W^\intercal K)R]^\intercal$
  $H = H * P/(H\text{crossprod}(W))$
  $P = SH + K(RH)$
  $W = W * P/(W\text{crossprod}(H))$
**end**

---

Overall, note that the data-intensive computations on $T$ in these algorithms are all expressed as vectorized LA operations, as underscored in Section 3.1. Thus, our normalized matrix abstraction and rewrite rules enable MORPHEUS to automatically factorize all these algorithms in a unified way.

## 5. EXPERIMENTS

We compare the runtime performance of our rewrite rules for key LA operators and the four automatically factorized ML algorithms. Our goal is to evaluate the speed-ups provided by MORPHEUS and understand how they vary for different data dimensions. Both synthetic and real-world datasets are used.

**Datasets.** We generate synthetic datasets for PK-FK and M:N joins with a wide range of data dimensions as listed in Table 4 and Table 5, respectively. For the PK-FK joins, the quantities varied are the *tuple ratio* ($n_S/n_R$) and *feature ratio* ($d_R/d_S$), which as explained in [26], help quantify the amount of redundancy introduced by a PK-FK join. The other parameters are fixed as per Table 4. For M:N joins, the number of tuples, number of features, and join attribute domain size are varied and the other parameters fixed as per Table 5. Seven real-world normalized datasets are adapted from [28] for the ML algorithms. These datasets are represented as sparse feature matrices to handle nominal features. Recall that MORPHEUS supports both dense and sparse matrices. The dimensions and sparsity are listed in Table 6. All real datasets have numeric *target* features in **S**, which we binarize for logistic regression and treat as regular features for K-Means and GNMF. The schemas and features are listed in the appendix.

Table 4: Data dimension parameters for PK-FK joins.

| PK-FK Join | $n_S$ | $d_S$ | $n_R$ | $d_R$ |
|---|---|---|---|---|
| Tuple Ratio | Varied | 20 | $10^6$ | 40 or 80 |
| Feature Ratio | $2 \times 10^7$ or $10^7$ | 20 | $10^6$ | Varied |

Table 5: Data dimension parameters for M:N joins. $n_U$ is the domain size (number of unique values) of $J_S/J_R$.

| M:N Join | $n_S = n_R$ | $d_S = d_R$ | $n_U$ |
|---|---|---|---|
| # Tuples | Varied | 200 or 100 | 1000 |
| # Features | $2 \times 10^5$ or $10^5$ | Varied | 1000 |
| Domain Size | $2 \times 10^5$ or $10^5$ | 200 | Varied |

Table 6: Dataset statistics for the real-world datasets.

| Dataset | $(n_S, d_S, \text{nnz})$ | $q$ | $(n_{R_i}, d_{R_i}, \text{nnz})$ |
|---|---|---|---|
| Expedia | 942142,27,5652852 | 2 | 11939,12013,107451 |
| | | | 37021,40242,555315 |
| Movies | 1000209,0,0 | 2 | 6040,9509,30200 |
| | | | 3706,3839,81532 |
| Yelp | 215879,0,0 | 2 | 11535,11706,380655 |
| | | | 43873,43900,307111 |
| Walmart | 421570,1,421570 | 2 | 2340,2387,23400 |
| | | | 45,53,135 |
| LastFM | 343747,0,0 | 2 | 4099,5019,39992 |
| | | | 50000,50233,250000 |
| Books | 253120,0,0 | 2 | 27876,28022,83628 |
| | | | 49972,53641,249860 |
| Flights | 66548,20,55301 | 3 | 540,718,3240 |
| | | | 3167,6464,22169 |
| | | | 3170,6467,22190 |

**Experimental Setup.** All experiments were run on a machine with 20 Intel Xeon E5-2660 2.6 GHz cores, 160 GB RAM, and 3 TB disk with Ubuntu 14.04 LTS as the OS. Our code is implemented in R v3.2.3 and uses the inbuilt default BLAS package libblas3 v1.2.20110419-7 . Since all real datasets fit in memory as R matrices, we use MORPHEUS on standard R for all experiments, except for the scalability study with MORPHEUS on ORE.

## 5.1 Operator-level Results

We first study the effects of the rewrites on individual LA operator runtimes using synthetic data. This will help us understand the runtime results for the ML algorithms later. The data preparation time is excluded for both the *materialized* version (in short, **M**), viz., joining the tables, and for the factorized version (in short, **F**), viz., constructing $K$ (or $I_S$ and $I_R$) matrices. As mentioned before, this pre-processing time was a minor fraction of the total runtimes in almost all cases on the real data. Furthermore, the time to construct the normalized matrix was almost always smaller than the time to materialize the single table. We present the pre-processing runtimes in the appendix.

**PK-FK Join.** Figure 3 shows the speed-ups of **F** over **M** for four key LA operators. Other operators exhibit similar trends and for brevity sake, we present their results in the appendix. Note that **F** is significantly faster than **M** for a wide range of data dimensions for all operators. The speed-ups increase with both the tuple ratio and feature ratio, but grow faster with the latter because the amount of redundancy in $T$, and thus, in the ML computations, increases faster with the feature ratio. Figure 3(b) shows that the speed-ups are slightly lower for LMM compared to scalar multiplication. This is because the rewrite rule for LMM has slightly higher overhead. Interestingly, Figures 3(c,d) show that the speed-ups for cross-product and pseudo-inverse grow much faster with the feature ratio. This is because their runtimes are at least quadratic in $d$, while the previous two operators have $O(d)$ runtimes.

**Heuristic Decision Rule.** Figure 3 also shows that **F** is indeed sometimes slower than **M**, as suggested earlier in Section 3.7. In these cases, the tuple ratios and/or feature ratios are very low. Since these regions exhibit an "L" shape, it motivates us to consider a heuristic decision rule that is a disjunctive predicate with two thresholds: if the tuple ratio is $< \tau$ or if the feature ratio is $< \rho$, we do not use **F**. We tune $\tau$ and $\rho$ *conservatively* using the speed-up results from all of our experiments on synthetic data; we set $\tau = 5$ and $\rho = 1$. This is conservative because it is unlikely to wrongly predict that a slow-down will not occur when it does, but it might wrongly predict that a slow-down will occur even though it does not; but even in the latter cases, the speed-ups of **F** over **M** were minor ($< 50\%$). We leave more sophisticated approaches to future work.

**M:N Join.** We now evaluate the rewritten operators for an M:N join. We set $(n_S, d_S) = (n_R, d_R)$ and vary $n_U$. Define the "join attribute uniqueness degree" as $n_U/n_S$. Note that as $n_U$ becomes smaller, more tuples are repeated after the join. $n_U = 1$ leads to the full cartesian product. Figure 4 presents the speed-ups for two key operators that arise in ML: LMM and cross-product. Other operators and other parameters are discussed in the appendix. We see that **F** is again significantly faster than **M** for a wide range of $n_U$ values for both operators. In fact, when $n_U = 0.01$, the speed-ups are nearly two orders of magnitude, which is comparable to the average number of times each tuple is repeated after the join. This confirms that our framework can efficiently handle M:N joins as well.

## 5.2 ML Algorithm-level Results

We compare the materialized versions (**M**) of the ML algorithms with the MORPHEUS-factorized versions (**F**) from Section 4. Due to space constraints, we focus primarily on a PK-FK join; M:N join is discussed in the appendix (the takeaways are similar). We study the effects of the data dimensions using synthetic data and then present the results on the real data. We then compare MORPHEUS against prior ML algorithm-specific factorized ML tools. Finally, we study the scalability of MORPHEUS on ORE.

### 5.2.1 Results on Synthetic Datasets

**Logistic Regression.** Figure 5(a) shows the runtimes for different tuple ratios and feature ratios, while the appendix presents a plot varying the number of iterations. It is clear that **F** is significantly faster than **M** in most cases across a wide range of data dimensions. The runtime is dominated by the LMM $Tw$ and the transposed LMM $T^\intercal P$ (which becomes an RMM). Thus, the speed-up trends are similar to that for those operators in Figure 3.

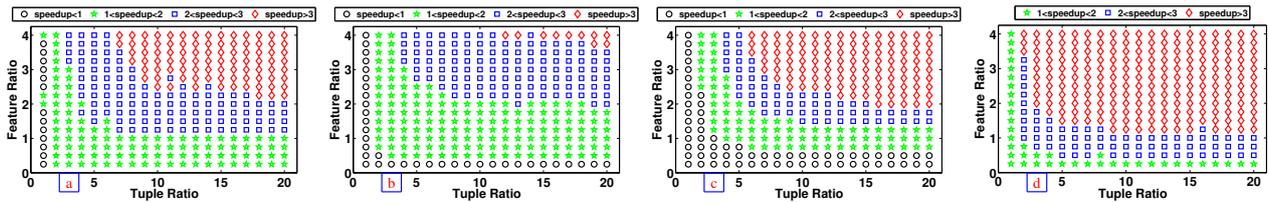

Figure 3: Speed-ups of factorized LA operators over the respective materialized versions on synthetic data for a PK-FK join. (a) Scalar multiplication, (b) LMM, (c) Cross-product, and (d) Pseudo-inverse. Data dimensions are in Table 4. For the cross-product, Algorithm 2 is used (a comparison with Algorithm 1 is presented in the appendix).

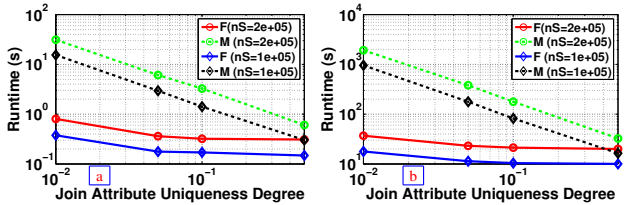

Figure 4: Runtimes for an M:N join of materialized (**M**) and factorized (**F**) versions of LA operators. (a) LMM and (b) Cross-product. We fix $n_S$ ($= n_R$) as shown on the plots, fix $d_S = d_R = 200$, and vary $n_U/n_S$ from 0.01 to 0.5.

Table 7: Runtimes (in seconds) on real data for the materialized approach (**M**) and speed-ups of MORPHEUS (**Sp**). E, M, Y, W, L, B, and F correspond to the datasets Expedia, Movies, Yelp, Walmart, LastFM, Books, and Flights, respectively. Number of iterations is 20 for all ML algorithms; number of centroids is 10 for K-Means, and number of topics is 5 for GNMF.

|   | Lin. Reg. | | Log. Reg. | | K-Means | | GNMF | |
|---|---|---|---|---|---|---|---|---|
|   | **M** | **Sp** | **M** | **Sp** | **M** | **Sp** | **M** | **Sp** |
| E | 73.1 | 22.2 | 71.2 | 14.0 | 102.7 | 4.5 | 80.9 | 5.9 |
| M | 20.3 | 36.3 | 65.4 | 30.3 | 93.3 | 6.0 | 75.4 | 8.0 |
| Y | 20.4 | 36.4 | 20.2 | 30.1 | 25.8 | 6.1 | 21.3 | 12 |
| W | 12.0 | 10.9 | 13.2 | 9.8 | 19.5 | 2.0 | 14.0 | 2.8 |
| L | 7.5 | 11.0 | 7.7 | 8.7 | 13.8 | 2.3 | 9.4 | 3.4 |
| B | 3.2 | 5.2 | 3.1 | 3.9 | 7.8 | 1.3 | 4.1 | 1.4 |
| F | 1.4 | 4.4 | 1.7 | 3.4 | 2.9 | 1.8 | 1.9 | 2.0 |

**Linear Regression.** The results are in Figure 5(b). Again, **F** is significantly faster than **M** for a wide range of data dimensions. The runtime is dominated by crossprod($T$) (see Algorithms 5 and 6). Thus, the speed-up trends are similar to that for crossprod in Figure 3. Gradient descent-based linear regression is similar to logistic regression; thus, we skip it for brevity and discuss it in the appendix.

**K-Means and GNMF.** The results are in Figure 5(c) and Figure 5(d), respectively. The runtime trends for the number of iterations for both are similar to logistic regression. Figure 5(c2) shows that K-Means runtime increases linearly with the number of centroids ($k$). The speed-up of **F** over **M** decreases slowly because unlike logistic regression, K-Means has extra computations beyond factorized operators whose contribution to the runtime increases as $k$ increases. The trends for GNMF are similar.

Table 8: Speed-ups of factorized logistic regression over materialized for a PK-FK join. Fix $(n_S, n_R, d_S, Iters) = (2 \times 10^6, 10^5, 20, 10)$; vary feature ratio ($d_R/d_S$).

| **Feature Ratio** | 1 | 2 | 3 | 4 |
|---|---|---|---|---|
| ORION [26] | 1.6 | 2.0 | 2.5 | 2.8 |
| MORPHEUS | 2.0 | 3.7 | 4.8 | 5.7 |

### 5.2.2 Results on Real Datasets

Since all the real datasets have multi-table star schema joins, this experiment also evaluates our multi-table join extension. Table 7 presents the results. We see that MORPHEUS is significantly faster than the materialized approach (**M**) in almost all cases for all datasets although the exact speed-ups differ widely across datasets and ML algorithms. The lowest speed-ups are mostly for GNMF and on Books, e.g., 1.4x for GNMF on Books, and 1.3x for K-Means on Books, primarily because the dataset has low feature ratios (as shown in Table 6) and GNMF and K-Means have extra computations after the factorized portions. On the other extreme, Movies and Yelp see the highest speed-ups, e.g., over 30x on both datasets for both linear regression and logistic regression. This is primarily because the runtimes of these ML algorithms are dominated by matrix multiplication operators, which are factorized by MORPHEUS, and these datasets have high feature and/or tuple ratios. Overall, these results validate MORPHEUS not only generalizes factorized ML, but also yields over an order of magnitude of speed-ups on some real datasets for a few popular ML tasks.

### 5.2.3 Comparison with ML Algorithm-specific Tools

We would like to know if the generality of MORPHEUS is at the cost of possible performance gains compared to prior ML algorithm-specific tools. Since the tool from [35] is not open sourced, we contacted the authors; after discussions, we realized that their tool does not support the equivalent of **M**, which makes an apples-to-apples comparison impossible. Thus, we only compare with the ORION tool from [26]. Note that ORION only supports dense features and PK-FK joins, unlike MORPHEUS. We vary the feature ratio and report the speed-ups in Table 8. MORPHEUS achieves comparable or higher speed-ups (in fact, the runtimes were also lower than ORION). This is primarily due to hashing overheads in ORION. Overall, we see that MORPHEUS provides high generality without sacrificing on possible performance gains.

### 5.2.4 Scalability with ORE

ORE executes LA operators over an *ore.frame* (physically, an RDBMS table) by pushing down computations to the RDBMS [2]. However, since ORE does not expose the underlying RDBMS multi-table abstractions (or the opti-mizer) to LA, by default, it needs the materialized single table. In contrast, MORPHEUS on ORE realizes the benefits of factorized ML on top of ORE (e.g., the *ore.rowapply* operator) without requiring changes to ORE. We compare the runtimes of MORPHEUS with the materialized version for logistic regression on larger-than-memory synthetic data. The results are presented in Table 9 (PK-FK join) and Table 10 (M:N join). We see that MORPHEUS yields speed-ups at scale for both PK-FK and M:N joins, validating our claim that MORPHEUS can leverage the scalability of existing LA systems. Since SystemML, SciDB, TensorFlow, and most other LA systems do not expose normalized data abstractions either, we expect our framework to benefit them too.

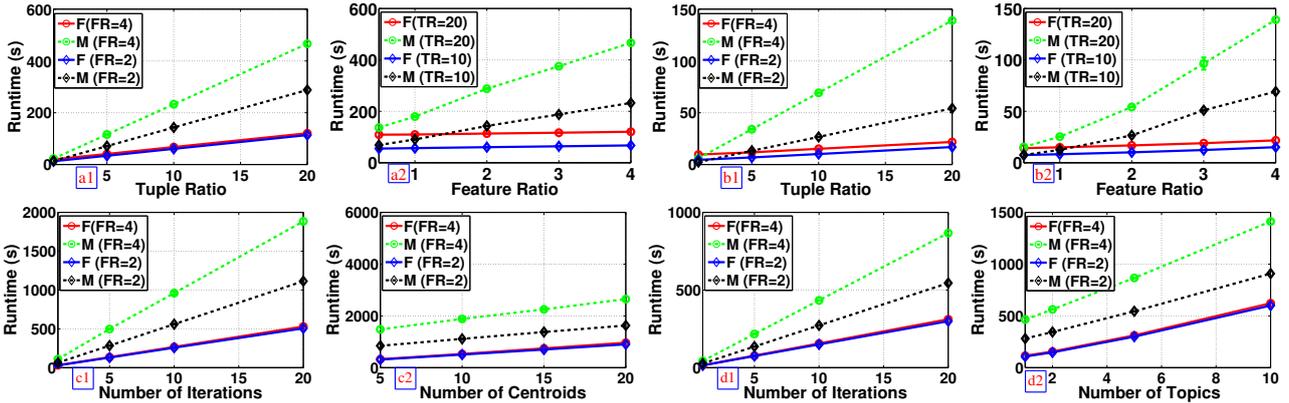

Figure 5: ML algorithms on synthetic data for a PK-FK join. Row 1: (a) Logistic Regression and (b) Linear Regression (using normal equations). Row 2: (c) K-Means and (d) GNMF. For (a), fix number of iterations to 20 (we vary this in the appendix. All data dimensions are listed in Table 4. For (c1) and (d1), we vary the number of iterations while fixing the number of centroids (resp. topics) for K-Means (resp. GNMF) to 10 (resp. 5). For (c2) and (d2), we set $(n_S, n_R, d_S) = (2 \times 10^7, 10^6, 20)$, while $d_R$ is 40 (FR=2) and 80 (FR=4), and number of iterations is 20 for both algorithms.

Table 9: Per-iteration runtime (in minutes) of logistic regression on ORE for a PK-FK join. Fix $(n_S, n_R, d_S) = (10^8, 5 \times 10^6, 60)$ and vary $d_R$ as per feature ratio (FR).

| Feature Ratio | 0.5 | 1 | 2 | 4 |
|---|---|---|---|---|
| Materialized | 98.27 | 130.09 | 169.36 | 277.52 |
| MORPHEUS | 56.30 | 62.51 | 68.54 | 73.33 |
| Speed-up | 1.8x | 2.1x | 2.5x | 3.8x |

Table 10: Per-iteration runtime (in minutes) of logistic regression on ORE for a M:N join. Fix $(n_S, n_R, d_S, d_R) = (10^6, 10^6, 200, 200)$ and vary $n_U$ (join attribute domain size).

| Domain Size | $5 \times 10^5$ | $10^5$ | $5 \times 10^4$ | $10^4$ |
|---|---|---|---|---|
| Materialized | 1.98 | 13.04 | 119.54 | 346.93 |
| MORPHEUS | 0.96 | 1.00 | 1.02 | 1.16 |
| Speed-up | 2.1x | 12.9x | 117.3x | 298.2x |

## 6. RELATED WORK

**Factorized ML.** As Figure 1(a) illustrates, prior works on factorized ML are either ML algorithm-specific or platform-specific or both [25, 26, 32, 34, 35]. For instance, [32, 35] only aim at optimizing linear regression, while [34] is restricted to in-memory datasets. Our work unifies and generalizes such ideas to a wider variety of ML algorithms, as well as data platforms. By factorizing LA operators, our work lets us decouple the ML algorithm from the platform, which enables us to leverage existing industrial-strength LA systems for scalability and other orthogonal benefits. It also lets data scientists automatically factorize any future ML algorithms expressible in LA systems.

**LA Systems.** Several tools support LA workloads over data systems [2, 4, 5, 9, 38]. There are also from-scratch systems for LA such as SciDB [14] and TensorFlow [5], both of which support tensor algebra, not just matrices. None of these systems optimize LA over normalized data. While they offer *physical data independence* for LA, our work brings *logical data independence* to LA. Since MORPHEUS offers closure, it could be integrated into any of these LA systems; our prototype on ORE is an encouraging first step in this direction. Related to our goals are two recent optimizations in SystemML: compressed LA (CLA) [18] and SPOOF [17]. CLA re-implements LA operators from scratch over compressed matrix formats to reduce memory footprints. MORPHEUS can be viewed as a schema-based form of compression. Unlike CLA, since MORPHEUS offers closure, it does not require re-implementing LA operators from scratch. Furthermore, CLA does not target runtime speed-ups [18] and thus, is complementary to MORPHEUS. SPOOF enables "sum-product" optimizations for LA expressions to avoid creating large intermediate matrices. While this is conceptually similar to avoiding join materialization, our work differs on both technical and architectural aspects. SPOOF does not exploit schema information, which means it cannot subsume factorized ML without an abstraction like our normalized matrix. Architecturally, SPOOF requires an LA compiler [17], while MORPHEUS also works in interpreted LA environments such as R and ORE. Overall, MORPHEUS is complementary to both CLA and SPOOF; it is interesting future work to integrate these ideas. To handle evolving data, LINVIEW proposed incremental maintenance for LA operators and expressions, albeit over single-table matrices [31]. To the best of our knowledge, most standard LA systems do not yet support incremental maintenance. Our work is orthogonal to this issue and it is interesting future work to integrate MORPHEUS and LINVIEW. Finally, BLAS is a popular library of fast low-level implementations of LA operations; it is the basic building block of many LA systems [30] and is used by several follow-on projects [15], including the widely used Linear Algebra PACKage (LA-

PACK) [6]. Our focus is orthogonal to such lower-level LA implementation issues; since our framework offers closure, it can be integrated into any of these LA packages.

**Query Optimization.** Factorized computations generalize prior work on optimizing SQL aggregates over joins [12,37]. In particular, FDB ("factorized database") is an in-memory tool that factorizes and optimizes relational algebra (RA) operations over joins [7]. In contrast, our focus is on LA operations with the aim of automatically factorizing many ML algorithms. This raises a grander question of whether LA can be "subsumed" by RA and RDBMSs, which is a long standing debate that is perhaps yet to be settled [14]. Different systems take different paths: [9,14] build from-scratch systems without an RDBMS, while [2,38] aim to layer LA on top of an RDBMS even if they do not fully exploit the RDBMS optimizer for LA. Our work is orthogonal to this debate; MORPHEUS is applicable to both kinds of systems, easily integrates with existing LA sytems, provides closure with respect to LA, and crucially, does not force ML users to learn RA or SQL. Furthermore, our work shows the benefits of database-style optimization ideas for LA operations regardless of the system environment, while also introducing new LA-specific optimizations with no known counterparts in RA. Nevertheless, it is interesting future work to more deeply integrate RA and LA, say, by creating a new representation language as suggested in [29]. There is also a need for benchmarks of LA systems in the spirit of [11]. While these questions are beyond the scope of this paper, such efforts could expose interesting new interactions between LA operations and optimizations such as multi-query optimization [27,36] and matrix chain product optimization [23] (implemented in Matlab [1] and SystemML [9]) coupled with join order optimization for normalized matrices.

## 7. CONCLUSION AND FUTURE WORK

Factorized ML techniques help improve ML performance over normalized data. But they have hitherto been ad-hoc and ML algorithm-specific, which causes a daunting development overhead when applying such ideas to other ML algorithms. Our work takes a major step towards mitigating this overhead by leveraging linear algebra (LA) to represent ML algorithms and factorizing LA. Our framework, MORPHEUS, generically and automatically factorizes several popular ML algorithms, provides significant performance gains, and can leverage existing LA systems for scalability. As ML-based analytics grows in importance, our work lays a foundation for more research on integrating LA systems with RA operations, as well as a grand unification of LA and RA operations and systems. As for future work, we are working on distributed versions of MORPHEUS on SystemML and TensorFlow. Another avenue is to include more complex LA operations such as Cholesky decomposition and SVD.

## Acknowledgements

This work was supported in part by gifts from Google and Microsoft, including a Google Faculty Research Award. We thank Matthias Boehm, Johann-Christoph Freytag, and the members of Wisconsin's Database Group and UC San Diego's Database Lab for their feedback. We thank Dan Olteanu and Maximilian Schleich for discussions on related work.

# APPENDIX

## A. OPERATOR REWRITES WITH TRANSPOSE

When a normalized matrix is transposed, i.e., we compute $T^\intercal$, the redundancy in $T$ is preserved but its structure changes. Optimizing a composition of transpose and other operators on a normalized matrix becomes a language level optimization, which makes the integration more complicated. For example, we might need to build a parser for R. As mentioned in Section 3.2, this issue is circumvented by adding a *transpose flag* to the normalized matrix data structure. This flag is set when $T$ is transposed and unset if it is transposed again. We now present a new set of rewrite rules that replace an operation on $T^\intercal$ with an operation on $T$, which means the rewrite rules from Section 3.3 can be reused.

**Element-wise Scalar Operators.** The output is a transposed normalized matrix.

$$T^\intercal \oslash x \rightarrow (T \oslash x)^\intercal \; ; \; x \oslash T^\intercal \rightarrow (x \oslash T)^\intercal$$
$$f(T^\intercal) \rightarrow (f(T))^\intercal$$

**Aggregation Operators.** The output is a column vector, row vector, or a scalar.

$$\text{colSums}(T^\intercal) \rightarrow (\text{rowSums}(T))^\intercal$$
$$\text{rowSums}(T^\intercal) \rightarrow (\text{colSums}(T))^\intercal \; ; \; \text{sum}(T^\intercal) \rightarrow \text{sum}(T)$$

**LMM and RMM.** The output is a regular matrix.

$$T^\intercal X \rightarrow (X^\intercal T)^\intercal$$
$$XT^\intercal \rightarrow (TX^\intercal)^\intercal$$

**Cross-product.** If the input is a transposed normalized matrix, this is the Gram matrix, which is used in some ML algorithms such as kernel-based SVMs. The output is a regular matrix.

$$\text{crossprod}(T^\intercal) \rightarrow \text{crossprod}(S^\intercal) + K \text{crossprod}(R^\intercal) K^\intercal$$

**Matrix Inversion Operators.** It suffices to use the rewrites devised for crossprod, LMM, RMM, and transposes directly:

$$ginv(T^\intercal) \rightarrow ginv(\text{crossprod}(T^\intercal))T$$
$$ginv(T^\intercal) \rightarrow T ginv(\text{crossprod}(T))$$

## B. DISCUSSION ABOUT MATRIX SOLVE

In practice, we have observed that the number of features is typically much less than that of tuples. Thus, the materialized matrix $T$ is often not square and therefore not invertible. When $T$ is indeed square, we have the following theorem:

THEOREM B.1. *Consider* $T_{n_S \times (d_S + d_R)} = [S_{n_S \times d_S}, K_{n_S \times n_r} \times R_{n_R \times d_R}]$. *If $T$ is invertable, then*

$$TR \leq \frac{1}{FR} + 1, \quad (1)$$

*where* $TR = \frac{n_S}{n_r}$ *and* $FR = \frac{d_R}{d_S}$.

PROOF. Invertibility of T implies that $KR$ is full column rank, which imples $R$ is full column rank and thus $d_R \leq n_R$. Noting that $T$ is square, we have

$$n_S = d_S + d_R = d_R(\frac{1}{FR} + 1) \leq n_R(\frac{1}{FR} + 1). \quad (2)$$

Moving $n_R$ to the left side, we obtain the result $TR \leq \frac{1}{FR} + 1$ and thus complete the proof. □

The theorem above indicates that invertibility of $T$ implies low redundancy.

## C. DOUBLE MATRIX MULTIPLICATION (DMM)

This is our name for the operation of multiplying two normalized matrices. While this scenario is rare in ML over a two-table join, it arises over multi-table joins (more in Section 3.5). Thus, we need to be able to rewrite this operation. Let the left normalized matrix be denoted $A = (S_A, K_A, R_A)$ and the right, $B = (S_B, K_B, R_B)$. Let $S_{B,1}$ and $S_{B,2}$ denote $S_B[1:d_{S_A},]$ and $S_B[(d_{S_A}+1):,]$ respectively. Similarly, let $K_{B,1}$ and $K_{B,2}$ denote $K_B[1:d_{S_A},]$ and $K_B[(d_{S_A}+1):,]$ respectively. Note that $d_A = n_B$. The rewrite is as follows:

$$AB \rightarrow [S_A S_{B,1} + K_A(R_A S_{B,2}),$$
$$(S_A K_{B,1})R_B + K_A((R_A K_{B,2})R_B)]$$

**Transposed DMM.** First, we consider both normalized matrix inputs ($A$ and $B$) being transposed. The output is a regular matrix.

$$A^\intercal B^\intercal \rightarrow (BA)^\intercal$$

Now, we consider a normalized matrix multiplied with a transposed normalized matrix (and vice versa). These are generalizations of the Gram Matrix and Gramian matrices. We are given two normalized matrices $A$ and $B$, similar to the case of double matrix multiplication. We need to rewrite $AB^\intercal$ and $A^\intercal B$, both of which yield regular matrices as outputs. For $AB^\intercal$, there are three cases: (1) $d_{S_A} = d_{S_B}$, (2) $d_{S_A} < d_{S_B}$, and (3) $d_{S_A} > d_{S_B}$. The rewrite for case (1) is as follows:

$$AB^\intercal \rightarrow S_A S_B^\intercal + K_A(R_A R_B^\intercal)K_B^\intercal$$

As for case (2), let $S_{B,1} = S_B[,1:d_{S_A}], S_{B,2} = S_B[,d_{S_A}+1:d_{S_B}], R_{A,1} = R_A[,1:d_{S_B}-d_{S_A}]$, and $R_{A,2} = R_A[,d_{S_B}-d_{S_A}+1:d_{R_A}]$. Note that we have $d_{S_A} + d_{R_A} = d_{S_B} + d_{R_B}$. The rewrite for case (2) is as follows:

$$AB^\intercal \rightarrow S_A S_{B,1}^\intercal + K_A(R_{A,1} S_{B,2}^\intercal) + K_A(R_{A,2} R_B^\intercal)K_B^\intercal$$

Finally, case (3) can be recast as case (2) with a transpose:

$$AB^\intercal \rightarrow (BA^\intercal)^\intercal$$

As for $A^\intercal B$, there is only one case but the rewrite is more complex.

$$A^\intercal B \rightarrow \begin{bmatrix} S_A^\intercal S_B & (S_A^\intercal K_B)R_B \\ R_A^\intercal(K_A^\intercal S_B) & R_A^\intercal K_A^\intercal K_B R_B \end{bmatrix}$$

An interesting thing to note in the above rewrite is that it is not straightforward to determine what the order of matrix multiplication should be for the fourth tile, viz., $R_A^\intercal K_A^\intercal K_B R_B$. If we compute the product $K_B R_B$ first, we are effectively materializing $B$. Thus, we might need to compute the product $K_A^\intercal K_B = P$ (say) first but this temporary output matrix could become a dense $n_{R_A} \times n_{R_B}$ matrix, which could be quite large depending on these dimensions. This requires a deeper understanding of the sparsity of $P$. We provide the following results that bound $\text{nnz}(P)$ from both below and above.

THEOREM C.1. $\text{nnz}(P) \geq \max\{n_{R_A}, n_{R_B}\}$

PROOF. Let $A_{i,\cdot}$, $A_{\cdot,j}$, and $A_{i,j}$ denote the $i^{th}$ row, the $j^{th}$ column, and the entry in the $i^{th}$ row and $j^{th}$ column of matrix $A$, respectively.

First, we prove by contradiction that $\text{nnz}(P_{i,\cdot}) \geq 1, \forall i = 1\ to\ n_{R_A}$, i.e., there is at least one non-zero element in the $i^{th}$ row of $P$. Suppose on the contrary that $\exists i\ s.t\ \text{nnz}(P_{i,\cdot}) = 0$. Note that $\text{nnz}(K_{A\cdot,i}) > 0$ implies that $\exists k\ s.t\ K_{Ak,i} = 1$. Thus, $\forall j = 1\ to\ n_{R_B}$, we have:

$$\begin{aligned} 0 &= P_{i,j} \\ &= K_{A\cdot,i}^T K_{B\cdot,j} \\ &= \sum_{w=1}^{n_{S_A}} K_{Aw,i} K_{Bw,j} \geq K_{Ak,i} K_{Bk,j} \\ &= K_{Bk,j} \end{aligned}$$

This implies that $\text{nnz}(K_{Bk,\cdot}) = 0$, which is a contradiction. Thus, $\text{nnz}(P_{i,\cdot}) \geq 1$. Therefore, we have:

$$\text{nnz}(P) = \sum_{i=1}^{n_{R_A}} \text{nnz}(P_{i,\cdot}) \geq n_{R_A}$$

Similarly, it can be shown that $\text{nnz}(P) \geq n_{R_B}$. Thus, $\text{nnz}(P) \geq \max\{n_{R_A}, n_{R_B}\}$. □

THEOREM C.2. $\text{nnz}(P) \leq n_{S_A} (= n_{S_B})$.

PROOF. Note that the sum of all elements in $P$ is

$$\begin{aligned} \text{sum}(P) &= 1_{n_{R_A} \times 1}^T \cdot P \cdot 1_{n_{R_B} \times 1} \\ &= 1_{n_{R_A} \times 1}^T \cdot K_A^T \cdot K_B \cdot 1_{n_{R_B} \times 1} \\ &= (K_A \cdot 1_{n_{R_A} \times 1})^T \cdot (K_B \cdot 1_{n_{R_B} \times 1}) \\ &= (1_{n_{S_A} \times 1})^T \cdot (1_{n_{S_B} \times 1}) \\ &= n_{S_A} (= n_{S_B}). \end{aligned}$$

Since the entries of $P$ are all nonnegative integers, the number of non-zeros in $P$ is bounded by its sum, i.e., $\text{nnz}(P) \leq \text{sum}(P)$. Thus, $\text{nnz}(P) \leq n_{S_A} (= n_{S_B})$. □

Thus, even in the worst case, $\text{nnz}(P) = n_{S_A}$. Depending on how large this is realtive to $n_{R_A} n_{R_B}$, $P$ will be sparse or dense. It is an open question as to whether one can estimate $\text{nnz}(P)$ before computing the product. This is similar to the classical problem of cardinality estimation. We leave it to future work to apply ideas from that literature for this problem. In our implementation, we use the simple strategy of always computing $P$ first. An alternative hybrid strategy is to always materialize the fourth tile to avoid computing $P$, but use the factorized multiplication for the second and third tiles. It is worth noting that the degenerate case of $A = B$ makes this rewrite similar to the naive method for the cross-product Algorithm 1.

**Algorithm 9:** Cross-product for M:N join (Naive)

$P = R^\intercal((I_R^\intercal I_S)S)$
$T_S = (S^\intercal((I_S^\intercal I_S)S)$
$T_R = (R^\intercal((I_R^\intercal I_R)R)$
**return** $\begin{bmatrix} T_S & P^\intercal \\ P & T_R \end{bmatrix}$

**Algorithm 10:** Cross-product for M:N join (Efficient)

$P = R^\intercal((I_R^\intercal I_S)S)$
$T_S = \text{crossprod}\left(\text{diag}\left(\text{colSums}(I_S)\right)^{\frac{1}{2}} S\right)$
$T_R = \text{crossprod}\left(\text{diag}\left(\text{colSums}(I_R)\right)^{\frac{1}{2}} R\right)$
**return** $\begin{bmatrix} T_S & P^\intercal \\ P & T_R \end{bmatrix}$

## D. EXTENSION TO M:N JOINS

This section shows the extension of our framework to the rewrite rules for general M:N joins.

**Element-wise Scalar Operators.**

$$T \oslash x \rightarrow (I_S, S \oslash x, I_R, R \oslash x)$$
$$x \oslash T \rightarrow (I_S, x \oslash S, I_R, x \oslash R)\ ;\ f(T) \rightarrow (I_S, f(S), I_R, f(R))$$

**Aggregation Operators.**

$\text{rowSums}(T) \rightarrow I_S \text{rowSums}(S) + I_R \text{rowSums}(R)$
$\text{colSums}(T) \rightarrow [\text{colSums}(I_S)S, \text{colSums}(I_R)R]$
$\text{sum}(T) \rightarrow \text{colSums}(I_S)\text{rowSums}(S) + \text{colSums}(I_R)\text{rowSums}(R)$

**LMM and RMM.**

$TX \rightarrow I_S(SX[1:d_S,])\ + I_R(RX[d_S+1:d_S+d_R,])$
$XT \rightarrow [(XI_S)S, (XI_R)R]$

**Cross-product.** The cross-product for M:N join using the naive method and the efficient method is presented in Algorithm 9 and Algorithm 10, respectively. On the other hand, the rewrite rule for the cross-product of $T^\intercal$ is as follows:

$\text{crossprod}(T^\intercal) \rightarrow I_S \text{crossprod}(S^\intercal)I_S^\intercal + I_R \text{crossprod}(R^\intercal)I_R^\intercal$

Observe that if the join is PK-FK, we have $I_S = I$ (identity matrix of size $n_S \times n_S$) and the above rules implicitly become equivalent to their counterparts from Section 3.3.

## E. EXTRA REWRITE RULES FOR MULTI-TABLE M:N JOINS

This section shows the rewrite rules for multi-table M:N joins. Consider the following joins:

$$T = R_1 \bowtie R_2 \bowtie \cdots \bowtie R_q, \tag{3}$$

where the join conditions are

$$R_j.JC_j = R_{J_j}.OJC_{J_j}, j = 1, 2, \cdots, q. \tag{4}$$

In other words, each table $R_j$ is joined with another table $R_{J_j}$ on $R_j$'s column $JC_j$ and $R_{J_j}$'s column $OJC_{J_j}$. Attach attributes $NR_j$ to table $R_j$ to encode row numbers, i.e., $NR_j$ takes values from 1 to the number of rows in $R_j$, $n_{R_j}$. Let $J_{R_j} = \{R_j.C|C = JC_j or \exists k, OJC_{J_k} = C\}$ be all the join

attributes involved in the multi-table join in table $R_j$. Now compute

$$T' = \pi_{NR_1, J_{R_1}}(R_1) \bowtie \pi_{NR_2, J_{R_2}}(R_2) \\ \bowtie \cdots \bowtie \pi_{NR_q, J_{R_q}}(R_q) \quad (5)$$

where the join conditions are still

$$R_j.JC_j = R_{J_j}.OJC_{J_j}, j = 1, 2, \cdots, q. \quad (6)$$

Then the indicator matrices $I_{R_j}, j = 1, 2, \cdots, q$ can be computed by

$$[I_{R_j}]([i, k]) = \begin{cases} 1, & \text{if } i^{th} \text{ row of } \mathbf{T}'.[NR_j]) = k \\ 0, & \text{otherwise} \end{cases}$$

Note that $I_{R_j}$ is also very sparse since $\text{nnz}(I_{R_j}) == |T'|$. W.L.O.G, assume each column of $I_{R_j}$ has at least one non-zero, i.e., each tuple of $I_{R_j}$ contributes to at least one tuple of $\mathbf{T}$. otherwise, we can remove those tuples a priori. Note that the extended normalized matrix is $(I_{R_1}, I_{R_2}, \cdots, I_{R_q}, R_1, R_2, \cdots, R_q)$, and $T = [I_{R_1}R_1, I_{R_2}R_2, \cdots, I_{R_q}R_q]$.

**Element-wise Scalar Operators.**

$$T \oslash x \to (I_{R_1}, \ldots, I_{R_q}, R_1 \oslash x, \ldots, R_q \oslash x) \\ x \oslash T \to (x \oslash S, I_{R_1}, \ldots, I_{R_q}, x \oslash R_1, \ldots, x \oslash R_q), \\ f(T) \to (f(S), I_{R_1}, \ldots, I_{R_q}, f(R_1), \ldots, f(R_q)).$$

**Aggregation Operators.**

$$\text{rowSums}(T) \to \sum_{i=1}^{q} I_{R_i} \text{rowSums}(R_i)$$

$$\text{colSums}(T) \to [\text{colSums}(I_{R_1})R_1, \cdots, \text{colSums}(I_{R_q})R_q]$$

$$\text{sum}(T) \to \sum_{i=1}^{q} \text{colSums}(I_{R_i})\text{rowSums}(R_i)$$

**LMM and RMM.** Let the dimensions of $R_i$ be $n_{R_i} \times d_{R_i}$ and $d = \sum_{i=1}^{Q} d_{R_i}$. Define $d'_i = \sum_{j=1}^{i} d_{R_j}$, for $i = 1$ to $q$, and $d'_0 = 0$. Given $X$ of size $d \times m$ ($m \geq 1$), the rewrite rules are as follows.

$$TX \to \sum_{i=1}^{q} I_{R_i}(R_i X[d'_{i-1} + 1 : d'_i,])$$

$$XT \to [(XI_{R_1})R_1, \ldots, (XI_{R_q})R_q]$$

**Cross-product.** For the sake of readability, let $C_i$ and $\text{cp}(X)$ denote $I_{R_i}R_i$ and $\text{crossprod}(X)$, respectively. Using the block matrix multiplication, $\text{cp}(T) = T^{\intercal}T = [I_{R_1}R_1, \ldots, I_{R_q}R_q]^{\intercal}[I_1R_1, \ldots, I_{R_q}R_q]$ can be rewritten as follows.

$$\begin{bmatrix} \text{cp}(C_1) & C_1^{\intercal}C_2 & \cdots & C_1^{\intercal}C_q \\ \text{cp}(C_1) & C_1^{\intercal}C_2 & \cdots & C_1^{\intercal}C_q \\ C_2^{\intercal}C_1 & \text{cp}(C_2) & \cdots & C_2^{\intercal}C_q \\ \vdots & \vdots & \ddots & \vdots \\ C_q^{\intercal}C_1 & C_q^{\intercal}C_2 & \cdots & \text{cp}(C_q) \end{bmatrix}$$

Since $\text{cp}(T)$ is symmetric. we only need to compute the upper right parts of it, i.e., all diagonal block matrices $\text{cp}(C_i)$ and $C_i^{\intercal}C_j$. For each diagonal block matrix $\text{cp}(C_i)$, we use the rewrite rule $\text{crossprod}\left(\left(\text{diag}\left(\text{colSums}\left(I_{R_i}\right)\right)\right)^{\frac{1}{2}} R_i\right)$. For $C_i^{\intercal}C_j$, $R_i^{\intercal}(C_i^{\intercal}C_j)R_j$ is used.

---

**Algorithm 11:** Linear Regression/GD (Standard)
**Input:** Regular matrix $T, Y, w, \alpha$
**for** $i$ in $1 : max\_iter$ **do**
$\quad w = w - \alpha * (T^{\intercal}(Tw - Y))$
**end**

---

**Algorithm 12:** Linear Regression/GD (Factorized)
**Input:** Normalized matrix $(S, K, R), Y, w, \alpha$.
**for** $i$ in $1 : max\_iter$ **do**
$\quad P = (Sw[1 : d_S,] + K(Rw[d_S + 1 : d_S + d_R,])) - Y)^{\intercal}$
$\quad w = w - \alpha * [PS, (PK)R]^{\intercal}$
**end**

---

On the other hand, the rewrite rule for the cross-product of $T^{\intercal}$ is as follows:

$$\text{crossprod}(T^{\intercal}) \to \sum_{i=1}^{q} I_{R_i} \text{crossprod}(R_i^{\intercal}) I_{R_i}^{\intercal}.$$

## F. ASYMPTOTIC PERFORMANCE

The asymptotic performance of all operators for a PK-FK join is presented in Table 11.

## G. LINEAR REGRESSION WITH GD

The standard single-table version is presented in Algorithm 11, while the automatically factorized version is presented in Algorithm 12.

## H. LINEAR REGRESSION IN SCHLEICH ET AL. FROM SIGMOD'16

The standard single-table version of their hybrid algorithm for linear regression is presented in Algorithm 13, while the automatically factorized version is presented in Algorithm 14.

## I. STANDARD VERSIONS OF K-MEANS AND GNMF

Algorithm 15 presents the standard single-table version of K-Means clustering expressed in linear algebra. Algorithm 16 presents the standard single-table version of GNMF.

## J. SCHEMAS OF REAL DATASETS

The datasets are adapted from [28]. Essentially, we convereted categorial features into large sparse matrices, which means the feature matrices are sparse. Note that the foreign key is a feature too and it is folded into the attribute table feature matrix ($R$). The following is a description of the datasets and their schemas from [28].

*Walmart*. Predict department-wise sales by joining data about past sales with data about stores and weather/economic indicators. **S** is Sales (SalesLevel, IndicatorID, StoreID, Dept), while $Y$ is SalesLevel. **R**$_1$ is Indicators (<u>IndicatorID</u>, TempAvg, TempStdev, CPIAvg, CPIStdev, FuelPriceAvg, FuelPriceStdev, UnempRateAvg, UnempRateStdev, IsHoliday). **R**$_2$ is Stores (<u>StoreID</u>, Type, Size). Both foreign keys (StoreID and IndicatorID) have closed domains with respect to the prediction task.

| Operator | Standard | Factorized | $\lim_{TR \to +\infty} Sp$ | $\lim_{FR \to +\infty} Sp$ |
|---|---|---|---|---|
| Scalar Function | $n_S(d_S + d_R)$ | $n_S d_S + n_R d_R$ | $1 + FR$ | $TR$ |
| Aggregation | | | | |
| LMM | $d_X n_S(d_S + d_R)$ | $d_X(n_S d_S + n_R d_R)$ | | |
| RMM | $n_X n_S(d_S + d_R)$ | $n_X(n_S d_S + n_R d_R)$ | | |
| Cross-Product | $\frac{1}{2}(d_S + d_R)^2 n_S$ | $\frac{1}{2}d_S^2 n_S + \frac{1}{2}d_R^2 n_R + d_S d_R n_R$ | $(1 + FR)^2$ | |
| Pseudo Inverse ($n > d$) | $7n_S(d_S+d_R)^2 + 20(d_S+d_R)^3$ | $27(d_S+d_R)^3 + \frac{1}{2}d_S^2 n_S$ $+\frac{1}{2}d_R^2 n_R + d_S d_R n_R$ $+(d_S+d_R)(n_S d_S + n_R d_R)$ | $14\frac{(1+FR)^2}{2FR+3}$ | |
| Pseudo Inverse ($n \le d$) | $7n_S^2(d_S + d_R) + 20n_S^3$ | $27n_S^3 + \frac{1}{2}n_S^2 d_S + \frac{1}{2}n_R^2 d_R +$ $+(n_S)(n_S d_S + n_R d_R)$ | | $14\frac{TR^2}{1+TR}$ |

Table 11: Asympototic Performance of the rewrite rules. $Sp$ is the standard computations divided by the factorized computations. $TR \equiv \frac{n_S}{n_R}$ and $FR \equiv \frac{d_R}{d_S}$. For ginv, the economic SVD is used internally using the standard R-SVD algorithm [20].

**Algorithm 13:** Linear Regression; Schleich et al. [35]

**Input:** Regular matrix $T$, $Y$, $w$.
$C = \begin{bmatrix} Y^\intercal T \\ \text{crossprod}(T) \end{bmatrix}$ //Co-factor
**for** $i$ in $1 : max\_iter$ **do**
$\quad w = w - \alpha * (C^\intercal \begin{bmatrix} -1 \\ w \end{bmatrix})$ //AdaGrad
**end**

**Algorithm 14:** Linear Regression (Schleich et al. [35]); Factorized

**Input:** Normalized matrix $(S, K, R)$, $Y$, $w$.
$P_1 = [Y^\intercal S, (Y^\intercal K)R]$
$P_2 = \text{crossprod}((S, K, R))$ //Use Algo. 2
$C = \begin{bmatrix} P_1 \\ P_2 \end{bmatrix}$ //Co-factor
**for** $i$ in $1 : max\_iter$ **do**
$\quad w = w - \alpha * (C^\intercal \begin{bmatrix} -1 \\ w \end{bmatrix})$ //AdaGrad
**end**

**Algorithm 15:** K-Means Clustering (Standard)

**Input:** Regular matrix $T$, # centroids $k$
//Initialize centroids matrix $C_{d \times K}$
//$\mathbf{1}_{a \times b}$ represents an all 1 matrix in $\mathbb{R}^{a \times b}$; used for replicating a vector row-wise or column-wise
//1. Pre-compute $l^2$-norm of points for distances
$D_T = \text{rowSums}(T\textasciicircum 2)\mathbf{1}_{1 \times k}$
$T_2 = 2 \times T$
**for** $i$ in $1 : max\_iter$ **do**
$\quad$//2. Compute pairwise squared distances; $D^{n \times k}$ has points on rows and centroids/clusters on columns.
$\quad D = D_T - T_2 C + \mathbf{1}_{n \times 1}\text{colSums}(C\textasciicircum 2)$
$\quad$//3. Assign each point to nearest centroid; $A^{n \times k}$ is a boolean (0/1) assignment matrix
$\quad A = (D == (\text{rowMin}(D)\mathbf{1}_{1 \times k}))$
$\quad$//4. Compute new centroids; denominator counts number of points in the new clusters, while numerator adds up assigned points per cluster
$\quad C = (T^\intercal A)/(\mathbf{1}_{d \times 1}\text{colSums}(A))$
**end**

**Algorithm 16:** Gaussian NMF (Standard)

**Input:** Regular matrix $T$, rank $r$
//Initialize $W$ and $H$
**for** $i$ in $1 : max\_iter$ **do**
$\quad H = H * (T^\intercal W)/(H \text{crossprod}(W))$
$\quad W = W * (TH)/(W \text{crossprod}(H))$
**end**

*Expedia*. Predict if a hotel's rank by joining data about past search listings with data about hotels and search events. **S** is Listings (Position, HotelID, SearchID, Score1, Score2, LogHistoricalPrice, PriceUSD, PromoFlag, OrigDestDistance). $Y$ is Position. **R**$_1$ is Hotels (HotelID, Country, Stars, ReviewScore, BookingUSDAvg, BookingUSDStdev, BookingCount, BrandBool, ClickCount). **R**$_2$ is Searches (SearchID, Year, Month, WeekOfYear, TimeOfDay, VisitorCountry, SearchDest, LengthOfStay, ChildrenCount, AdultsCount, RoomCount, SiteID, BookingWindow, SatNightBool, RandomBool). SearchID does not have a closed domain and is not used as a feature.

*Flights*. Predict if a route is codeshared by joining data about routes with data about airlines, source, and destination airports. **S** is Routes (CodeShare, AirlineID, SrcAirportID, DestAirportID, Equipment1, ..., Equipment20). $Y$ is CodeShare. **R**$_1$ is Airlines (AirlineID, AirCountry, Active, NameWords, NameHasAir, NameHasAirlines). **R**$_2$ is SrcAirports (SrcAirportID, SrcCity, SrcCountry, SrcDST, SrcTimeZone, SrcLongitude, SrcLatitude). **R**$_3$ is DestAirports (DestAirportID, DestCity, DestCountry, DestTimeZone, DestDST, DestLongitude, DestLatitude).

*Yelp*. Predict business ratings by joining data about past ratings with data about users and businesses. **S** is Ratings (Stars, UserID, BusinessID). $Y$ is Stars. **R**$_1$ is Businesses (BusinessID, BusinessStars, BusinessReviewCount, Latitude, Longitude, City, State, WeekdayCheckins1, ..., WeekdayCheckins5, WeekendCheckins1, ..., WeekendCheckins5, Category1, ... Category15, IsOpen). **R**$_2$ is Users (UserID, Gender, UserStars, UserReviewCount, VotesUseful, VotesFunny, VotesCool).

*MovieLens1M*. Predict movie ratings by joining data about past ratings with data about users and movies. **S** is Ratings (Stars, UserID, MovieID). $Y$ is Stars. **R**$_1$ is Movies (MovieID, NameWords, NameHasParentheses, Year, Genre1,

..., Genre18). $\mathbf{R}_2$ is Users (UserID, Gender, Age, Zipcode, Occupation).

**LastFM**. Predict music play counts by joining data about past play levels with data about users and artists. $\mathbf{S}$ is Plays (PlayLevel, UserID, ArtistID), $Y$ is PlayLevel, $\mathbf{R}_1$ is Artists (ArtistID, Listens, Scrobbles, Genre1, ..., Genre5), and $\mathbf{R}_2$ is Users (UserID, Gender, Age, Country, JoinYear).

**BookCrossing**. Predict book ratings by joining data about past ratings with data about readers and books. $\mathbf{S}$ is Ratings (Stars, UserID, BookID). $Y$ is Stars. $\mathbf{R}_1$ is Users (UserID, Age, Country). $\mathbf{R}_2$ is Books (BookID, Year, Publisher, NumTitleWords, NumAuthorWords).

## K. DATA PREPARATION

The data preparation time for all real datasets is presented in Table 12. Since many ML algorithms are iterative, the one-time cost of constructing the normalized matrix is typically a small fraction of the overall runtime. The data in Table 12 verified the statement above, as for most of the case, the data preparation time accounted for at most 5% of the total runtime of Logistic Regression.

|   | Data. Prep. | | Log. Reg. | | Ratio | |
|---|---|---|---|---|---|---|
|   | M | F | M | F | M | F |
| E | 1.85 | 0.15 | 71.2 | 5.09 | 0.026 | 0.029 |
| M | 1.63 | 0.13 | 65.4 | 2.16 | 0.025 | 0.060 |
| Y | 0.49 | 0.03 | 20.2 | 0.67 | 0.024 | 0.045 |
| W | 0.51 | 0.05 | 13.2 | 1.35 | 0.039 | 0.037 |
| L | 0.32 | 0.05 | 7.7 | 0.89 | 0.042 | 0.056 |
| B | 0.28 | 0.03 | 3.1 | 0.79 | 0.090 | 0.038 |
| F | 0.16 | 0.01 | 1.7 | 0.50 | 0.094 | 0.020 |

Table 12: Runtimes (in seconds) for data preparation. Data. Prep. is the data preparation time. Log. Reg. is the runtime of logsitic regresion with 20 iterations. Ratio is preparation time divided by the logistic regression runtime.

## L. MORE RUNTIME PLOTS

The runtime and speed-up plots on synthetic data for scalar addition, RMM, column summation, and (full) summation are presented in Figure 6. The runtime plots on synthetic data for scalar multiplication, LMM, cross-product, and row summation are presented in Figure 7. The runtime plot on synthetic data for logistic regression for varying number of iterations is presented in Figure 9. The runtime plots on synthetic data for linear regression with gradient descent are presented in Figure 8. The runtime plots on synthetic data for varying tuple and feature ratios for K-Means and GNMF are presented in Figure 10.

## M. PLOTS FOR M:N JOIN

We vary the number of tuples, number of features, and join attribute uniqueness fraction ($n_U/n_S$) for various key LA operators. The results are presented in Figures 11 and 12.

## N. ORACLE R ENTERPRISE IMPLEMENTATION NOTE

Our system for Oracle R Enterprise (ORE) is built on top of ore.rowapply, a function in ORE to deal with larger-than-memory data. It partitions a large table(matrix) and performs operations specified by users to each chunk of the table, and then returns the output. This function is used to build LA operators (such matrix multiplications) for larger-than-memory data. Since our rewrite rules provide closure with respect to LA operators, our rewrite rules can be directly applied as long as the LA operators are available.

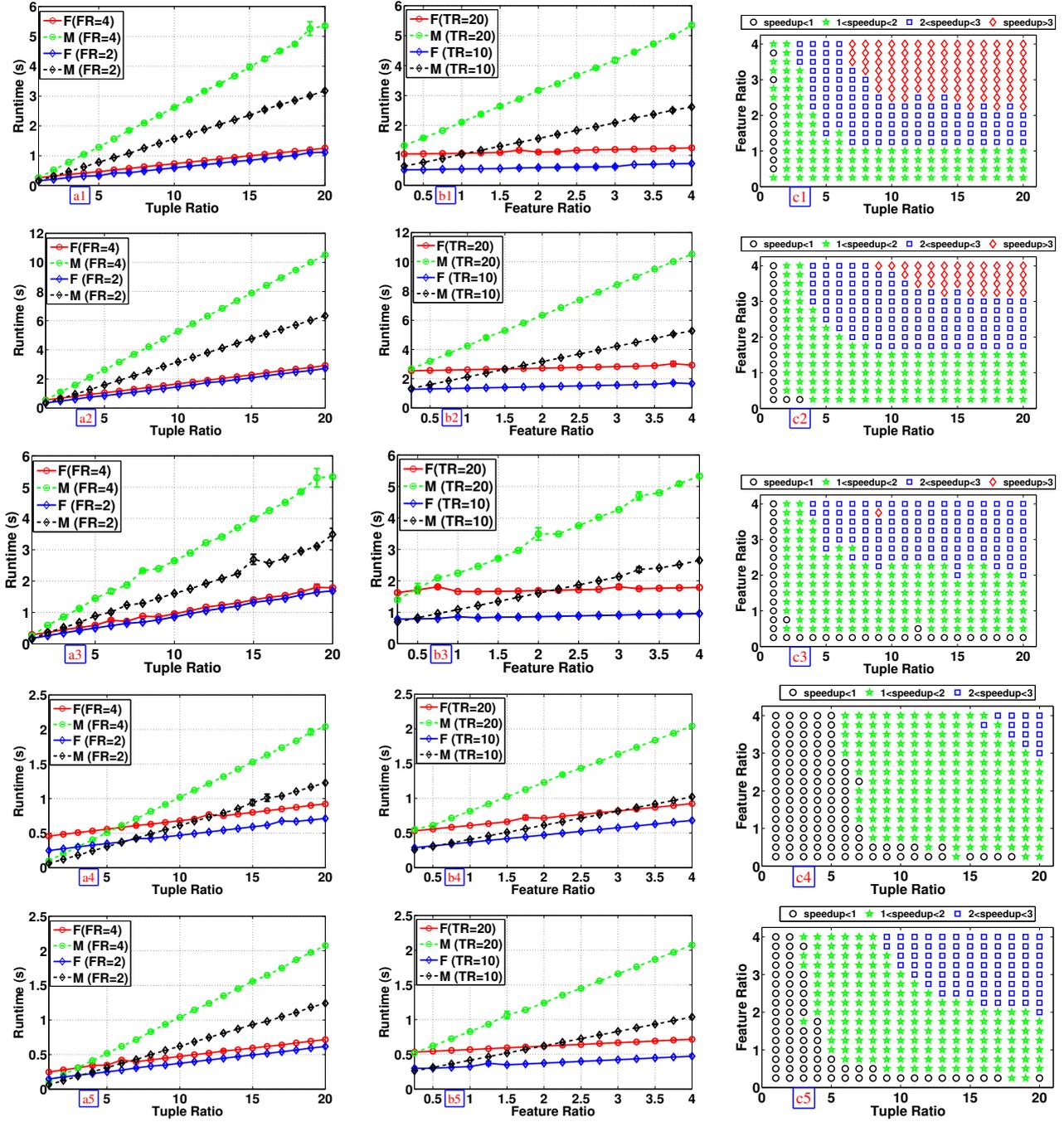

Figure 6: Runtimes for the same setting as Figure 3. The rows are: (1) Scalar addition, (2) RMM, (3) Row summation (4) Column summation, and (5) Summation. The columns are: (a) Vary tuple ratio (TR) $n_S/n_R$, (b) Vary feature ratio (FR) $d_R/d_S$, and (c) Speed-ups of F over M.

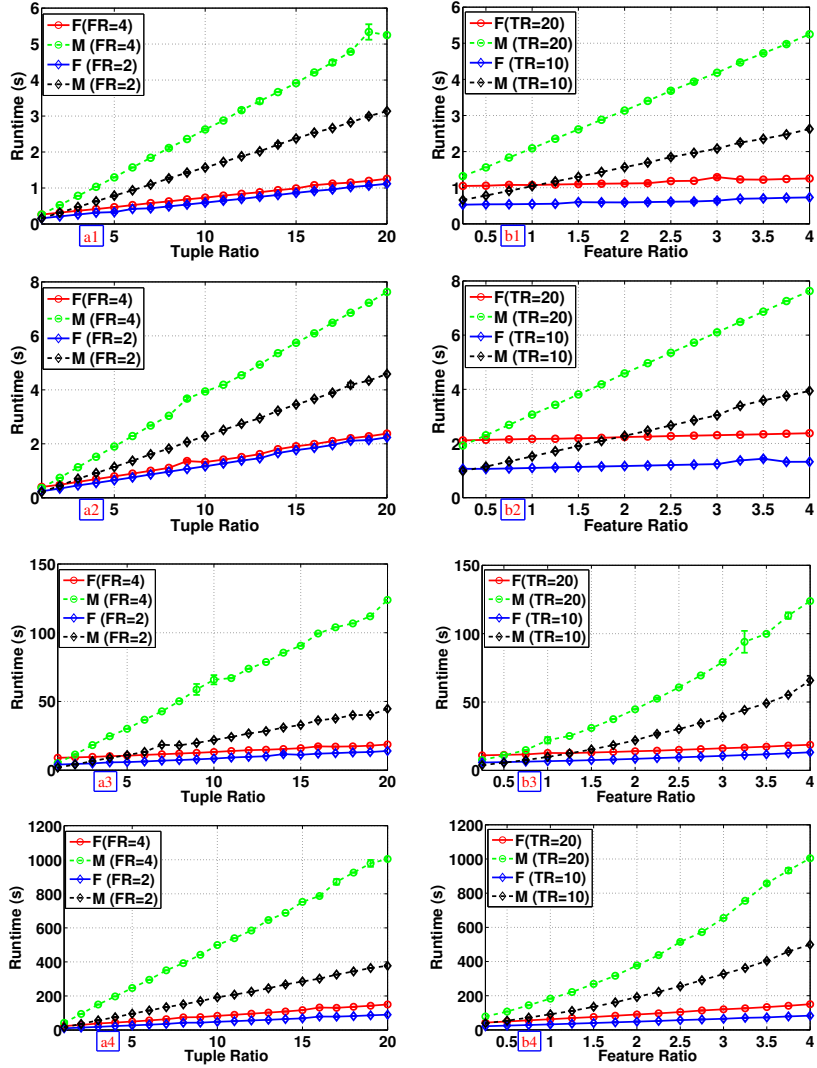

Figure 7: Runtimes for the same setting as Figure 3. The rows are: (1) Scalar multiplication, (2) LMM, (3) Cross-product, and (4) Pseudo-inverse. The columns are: (a) Vary tuple ratio (TR) $n_S/n_R$, and (b) Vary feature ratio (FR) $d_R/d_S$.

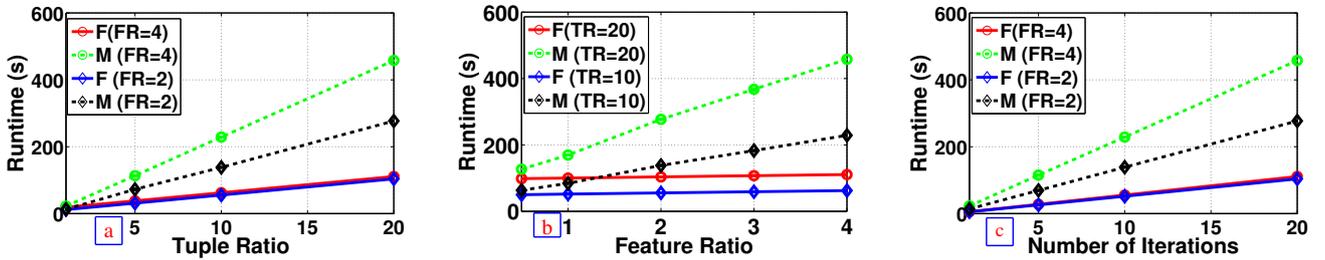

Figure 8: Linear Regression with gradient descent. (a) Vary TR. (b) Vary FR. (c) Vary the number of iterations. The data dimensions for (a) and (c) are listed in Table 4 and the number of iterations is 20. For (c), we fix $n_S = 2 \times 10^7$, $n_R = 10^6$, $d_S = 20$, and $d_R = 40$ (FR=2) and 80 (FR=4).

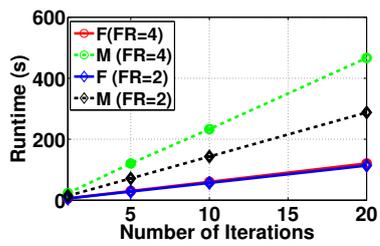

Figure 9: Logistic Regression for varying number of iterations. We set $(n_S, n_R, d_S) = (2 \times 10^7, 10^6, 20)$ and $d_R$ is 40 (FR=2) or 80 (FR=4).

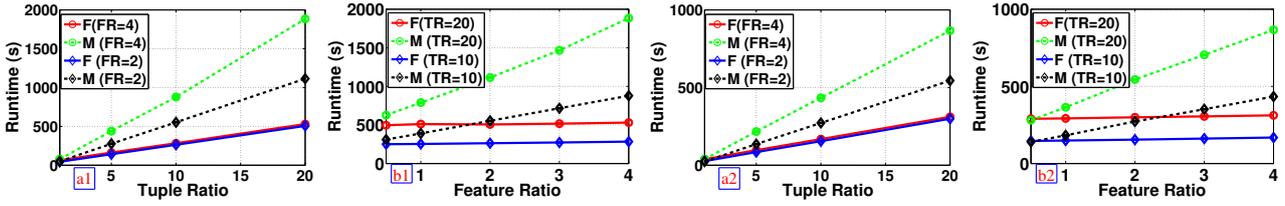

Figure 10: Runtimes on synthetic data. (1) K-Means and (2) GNMF. (a) Vary the tuple ratio and (b) Vary the feature ratio. The number of iterations is 20, $k = 10$ (for K-Means), and the number of topics is 5 (for GNMF). The data dimensions are listed in Table 4.

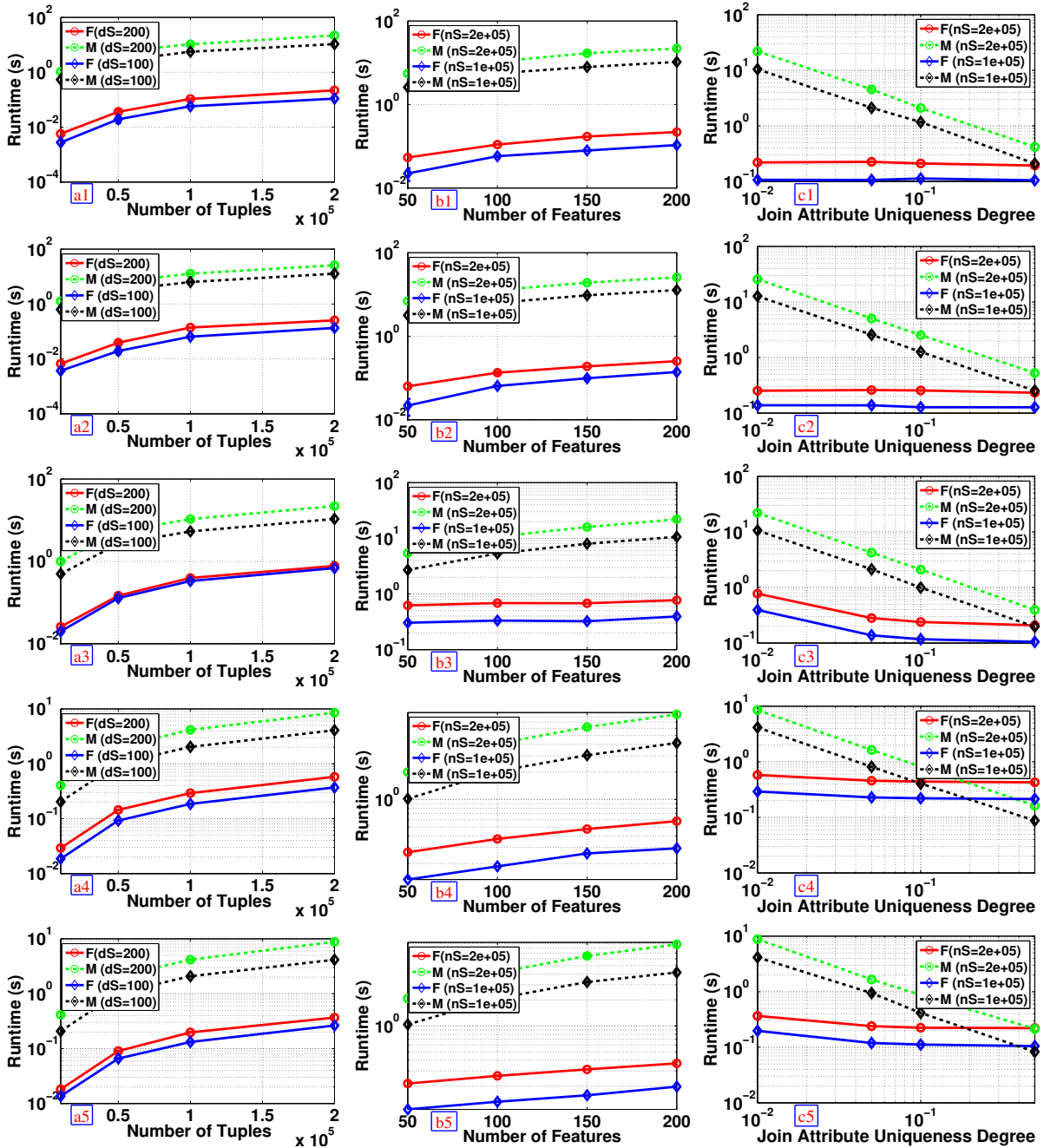

Figure 11: Operator-level results using synthetic data for M:N join. Row 1: Scalar Addition, Row 2: Scalar Multiplication, Row 3: Row Summation, Row 4: Column Summation, and Row 5: Summation.

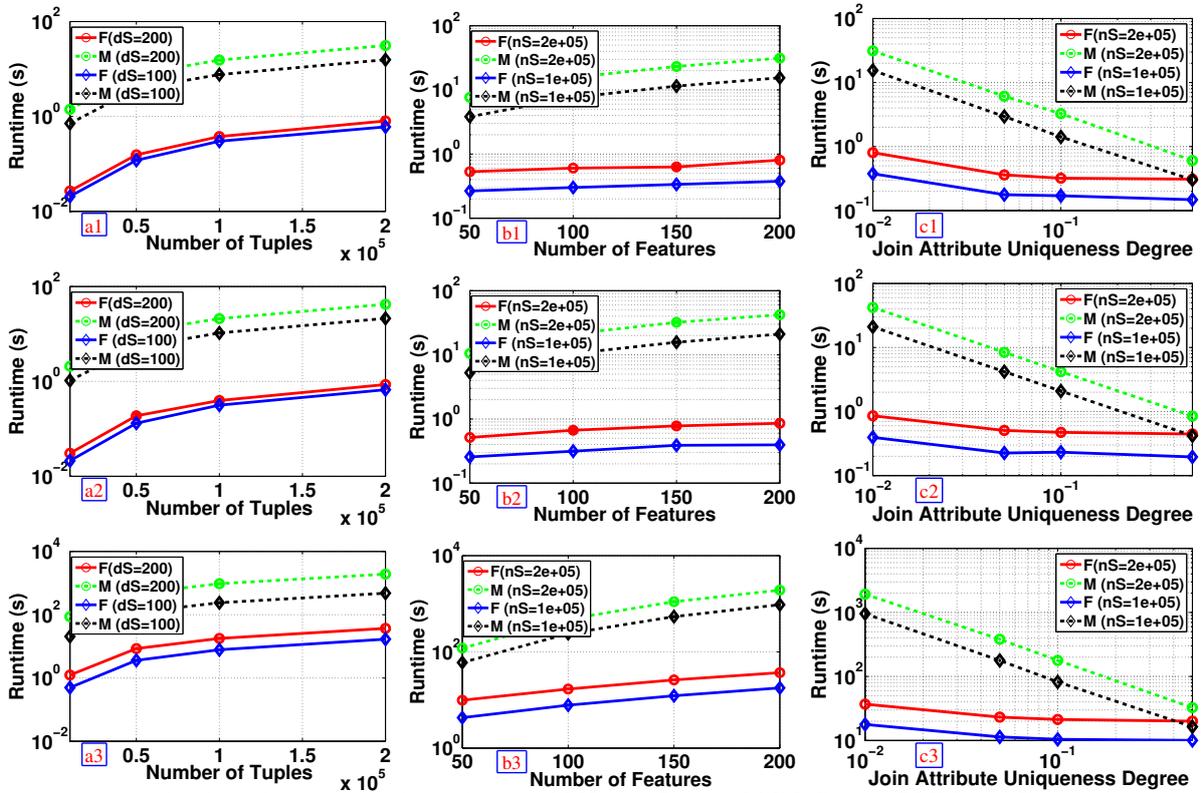

Figure 12: Operator-level results using synthetic data for M:N join. Row 1: LMM, Row 2: RMM, and Row 3: Cross-product.